\documentclass[prd,aps,twocolumn,tightenlines,notitlepage,nofootinbib,preprintnumbers,letterpaper,superscriptaddress]{revtex4-1} 

\pdfoutput=1
\usepackage{graphicx,mathtools,tabu}
\usepackage{epstopdf}
\usepackage{amsmath, bm}
\usepackage{amsfonts}
\usepackage[colorlinks=true,breaklinks=true]{hyperref}
\usepackage{accents}
\usepackage[figure, table]{hypcap}
\usepackage{amssymb}
\usepackage{appendix}
\usepackage{comment}
\usepackage{bbold}
\usepackage{xcolor}
\usepackage{color}
\usepackage{slashed}
\usepackage{subfigure}
\usepackage{setspace}
\usepackage{footnote}
\usepackage{multirow}
\usepackage{braket}
\usepackage[normalem]{ulem}
\usepackage[utf8]{inputenc}
\usepackage{mathrsfs}
\usepackage{longtable}
\usepackage{enumitem}

\hypersetup{colorlinks,citecolor= nicegreen,linkcolor= nicered}
\definecolor{nicered}{rgb}{0.7,0.1,0.1}
\definecolor{nicegreen}{rgb}{0.1,0.5,0.1}

\LTcapwidth=\textwidth
\usepackage{url}
\usepackage{wrapfig}
\usepackage{braket}
\def\be   {\begin{equation}}  
\def\ee   {\end{equation}}
\def\ba   {\begin{array}}     
\def\ea   {\end{array}}
\def\bea  {\begin{eqnarray}}  
\def\eea  {\end{eqnarray}}
\def\bean {\begin{eqnarray*}}  
\def\eean {\end{eqnarray*}}

\def\to {\rightarrow}

\def\dm{\delta m^2}

\newcommand{\eV}{{\rm\ eV}}

\newcommand{\MeV}{{\rm\ MeV}}

\newcommand{\kpc}{{\rm\ kpc}}

\newcommand*{\pbar}[1]{\accentset{(-)}{#1}}

\graphicspath{{figures/}}

\begin{document}

\preprint{FERMILAB-PUB-21-225-T, NUHEP-TH/21-05, N3AS-21-009} 

\title{SN1987A still shining: A Quest for Pseudo-Dirac Neutrinos}
\author{Ivan Martinez-Soler}
\email{ivan.martinezsoler@northwestern.edu}
\affiliation{Northwestern University, Department of Physics \& Astronomy, 2145 Sheridan Road, Evanston, IL 60208, USA}
\affiliation{Theory Department, Fermi National Accelerator Laboratory, P.O. Box 500, Batavia, IL 60510, USA}
\affiliation{Colegio de F\'isica Fundamental e Interdisciplinaria de las Am\'ericas (COFI), 254 Norzagaray street, San Juan, Puerto Rico 00901.}
\author{Yuber F. Perez-Gonzalez}
\email{yfperezg@northwestern.edu}
\affiliation{Northwestern University, Department of Physics \& Astronomy, 2145 Sheridan Road, Evanston, IL 60208, USA}
\affiliation{Theory Department, Fermi National Accelerator Laboratory, P.O. Box 500, Batavia, IL 60510, USA}
\affiliation{Colegio de F\'isica Fundamental e Interdisciplinaria de las Am\'ericas (COFI), 254 Norzagaray street, San Juan, Puerto Rico 00901.}
\author{Manibrata Sen}
\email{manibrata@berkeley.edu}
\affiliation{Northwestern University, Department of Physics \& Astronomy, 2145 Sheridan Road, Evanston, IL 60208, USA}
\affiliation{Department of Physics, University of California Berkeley, Berkeley, California 94720, USA}

\begin{abstract}
Ever since the discovery of neutrinos, we have wondered if neutrinos are their own antiparticles. One remarkable possibility is that neutrinos have a pseudo-Dirac nature, predicting a tiny mass difference between active and sterile states. We analyze the neutrino data from SN1987A in the light of active-sterile oscillations and find a mild preference ($\Delta\chi^2\approx 3$) for $\dm=6.31\times 10^{-20}\eV^2$. Notably, the same data is able to exclude $\dm\sim[2.55,3.01]\times 10^{-20}\eV^2$ with $\Delta\chi^2> 9$, the tiniest mass differences constrained so far. We further consider the next-generation of experiments and demonstrate their sensitivity exploring the nature of the neutrino mass.
\end{abstract}

\maketitle
%%%%%%%%
%%%%%%%%%
\section{Introduction}
%%%%%%%%
%%%%%%%%%
The quest to understand the fundamental nature of neutrinos still remains the Holy Grail of neutrino physics. The general consensus is that neutrinos can either be Dirac or Majorana, depending on whether the net lepton number is a conserved symmetry of the Standard Model (SM) or not. Our inability to distinguish between the two rests on the fact that in the ultra-relativistic limit, Dirac and Majorana neutrinos behave identically in all experiments~\cite{Kayser:1981nw}. Hence, one needs to probe lepton-number-violation~\cite{Furry:1938zz}, search for non-relativistic neutrinos~\cite{Long:2014zva,Berryman:2018qxn,Millar:2018hkv}, or other kinds of non-standard neutrino physics to answer this crucial question~\cite{Balantekin:2018ukw,Funcke:2019grs,deGouvea:2019goq,deGouvea:2021ual}. 

However, there remains a possibility that Nature solves this dichotomy by preferring a middle ground, where neutrinos are Majorana, but they behave almost as if they were Dirac. The hypothesis that neutrinos are \emph{pseudo-Dirac} (PD) requires soft lepton-number violation, thereby introducing a tiny mass-splitting in the chiral components of the mass-eigenstates~\cite{Wolfenstein:1981kw,Petcov:1982ya,Bilenky:1983wt,Kobayashi:2000md,Anamiati:2017rxw,deGouvea:2009fp,Vissani:2015pss}. This allows for a \emph{fifty-fifty} admixture of active and sterile states, with a possible oscillation between the two governed by their tiny mass-squared differences $\delta m^2$. Note that such oscillations between active-sterile neutrinos, driven by a tiny mass-squared difference can also arise in other scenarios, for e.g., mirror models~\cite{Berezinsky:2002fa}.
%%%%%%%%%
%%%%%%%%%
\begin{figure}[!h]
\centering
\includegraphics[width=.48\textwidth]{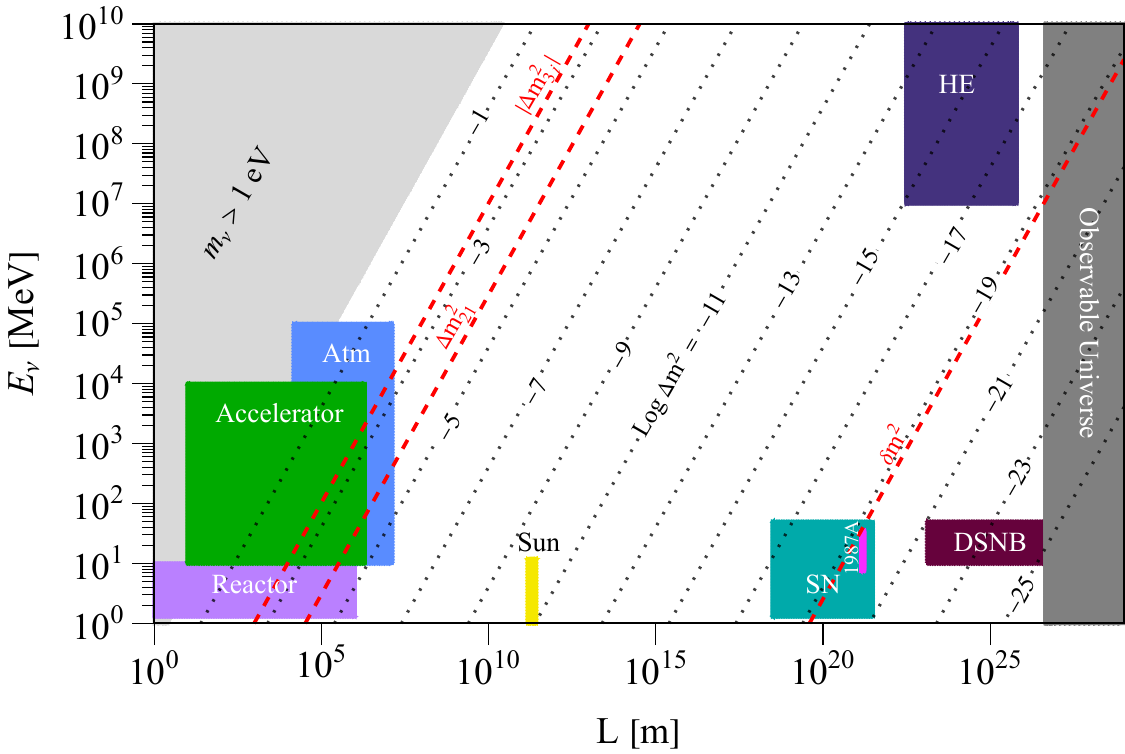}
\caption{Characteristic energies and baselines of distinct experiments with reactor (lilac), accelerator (green), atmospheric (light blue), solar (yellow), SN (emerald), DSNB (purple) and High Energy (violet) neutrinos. Dotted lines indicate the sensitivity to $\Delta m^2$ via \emph{vacuum} oscillations; we show three specific values in red for $|\Delta m_{3i}^2|,\Delta m_{21}^2, \dm$ , where, in the Normal Ordering, $\Delta m_{31}^2=2.51\times 10^{-3}\eV^2, \Delta m_{21}^2 = 7.42\times 10^{-5}\eV^2$~\cite{Esteban:2020cvm}, and $\dm = 6.31\times 10^{-20}\eV^2$. The particular case for SN1987A is highlighted with the fuchsia rectangle. \label{fig:PDbounds}}
\end{figure}
%%%%%%%%%
%%%%%%%%%

Testing this scenario is extremely difficult because these active-sterile oscillations take place over a baseline inversely proportional to the tiny mass-squared differences, so for all terrestrial experiments, these neutrinos behave like Dirac neutrinos. Since these oscillations can develop over astrophysical distances, the strongest bounds come from solar neutrinos ($\delta m^2\lesssim 10^{-12}\eV^2$)~\cite{deGouvea:2009fp}, and atmospheric neutrinos ($\delta m^2\lesssim 10^{-4}\eV^2$)~\cite{Beacom:2003eu}. Smaller values ($\delta m^2\sim 10^{-24}\eV^2$) can be tested with the measurement of the diffuse supernova background neutrinos (DSNB)~\cite{deGouvea:2020eqq}. Future terrestrial experiments, such as DUNE or JUNO, can also test the PD hypothesis, although they will be sensitive to larger quadratic mass differences~\cite{Anamiati:2019maf}. Collider signals of heavy PD neutrinos have also been explored~\cite{Das:2014jxa,Hernandez:2018cgc}. High energy astrophysical neutrinos would in principle be sensitive to mass difference of the order $10^{-18}\eV^2\lesssim \delta m^2\lesssim 10^{-12}\eV^2$~\cite{Beacom:2003eu,Keranen:2003xd,Esmaili:2009fk,Esmaili:2012ac,Joshipura:2013yba,Brdar:2018tce}. On the other hand, a galactic core-collapse supernova (SN), which releases almost its entire energy in the form of neutrinos, can provide the perfect astrophysical laboratory to test the PD nature of neutrinos. Such sensitivity, summarized in Fig.\,\ref{fig:PDbounds}, an updated version of the results presented in~\cite{Beacom:2003eu}, arises from the combination of the naturally long baseline and relatively small energy range of the emitted neutrinos, $E_\nu\sim{\cal O}(\rm MeV)$, and allows one to probe large oscillation lengths.

In this work, we utilize, for the first time, the SN1987A neutrino observation data from Kamiokande-II (KII)~\cite{Hirata:1987hu,Hirata:1988ad}, IMB~\cite{Bionta:1987qt,Bratton:1988ww}, and Baksan~\cite{Alekseev:1988gp} to probe the possible active-sterile oscillations in neutrinos over galactic-scale baselines. Performing an unbinned likelihood analysis, we find, quite intriguingly, that the combined data from the three experiments have a marginal preference for the PD scenario. Moreover, such a data allows for the exclusion of mass differences in the range $2.55\times 10^{-20}\eV^2 \lesssim \dm \lesssim 3.01\times 10^{-20}\eV^2$ with a $\Delta\chi^2>9$, the smallest values constrained yet. We further analyze the sensitivity of upcoming neutrino experiments like Hyper-Kamiokande and DUNE to utilise a future galactic SN, and find that, for a SN happening at 10\kpc, these experiments can probe values of $\delta m^2\sim [10^{-18},10^{-21}]\eV^2$. Clearly, a future galactic SN can allow us to probe such extreme values of the mass-squared differences which are not accessible to solar-system bound neutrino experiments. This, along with observations of the DSNB, as well as high energy neutrinos, can provide some of the most stringent bounds on the PD scenario.
%%%%%%%%%%%%%%%%%%%%%%%%%%%%%%%%%%%%%%%%%
%%%%%%%%%%%%%%%%%%%%%%%%%%%%%%%%%%%%%%%%%
%%%%%%%%
%%%%%%%%%
\section{Active-sterile oscillations}
%%%%%%%%
%%%%%%%%%
One of the most austere extensions of the SM to address neutrino masses consists of adding at least two right-handed neutrinos, singlets under the SM symmetries, and then implementing the Higgs mechanism. Nevertheless, gauge invariance allows for Majorana mass terms for the right-handed neutrinos. Thus, in general, the neutrino mass matrix below the electroweak scale is given by
\begin{align}\label{eq:MM}
    M_\nu=\begin{pmatrix}
        \mathbb{0}_3 & Y v/\sqrt{2}\\
        Y v/\sqrt{2} & M_R
    \end{pmatrix}\,,
\end{align}
$v/\sqrt{2}$ being the SM vacuum expectation value and $Y$ the Yukawa matrix. We have not considered heretofore any hierarchy in the mass matrix. The well-known see-saw mechanism \cite{Mohapatra:1979ia,GellMann:1980vs,Yanagida:1979as,Minkowski:1977sc,Mohapatra:1980yp,Magg:1980ut,Lazarides:1980nt,Wetterich:1981bx} assumes that the right-handed neutrino mass far exceeds the electroweak scale $M_{R}\gg Y v$, thus explaining the petiteness of neutrino masses.

On the other hand, if lepton number is softly broken, i.e., $M_{R}\ll Y v$, the small Majorana terms break the degeneracy between the masses of the left- and right-handed components, present in a purely Dirac neutrino. This can be an important scenario for neutrino masses, if experiments searching for lepton number violation return a null result. In such regime, the mass matrix $M_\nu$ can be diagonalized using the following unitary $6\times 6$ matrix $\mathscr{V}$~\cite{Kobayashi:2000md}
\begin{align}
    \mathscr{V}=\begin{pmatrix}
    U & 0\\
    0 & U_R
    \end{pmatrix}\cdot\frac{1}{\sqrt{2}}
    \begin{pmatrix}
    \mathbb{1}_3 & i \mathbb{1}_3\\
    \varphi & -i\varphi
    \end{pmatrix}\,,
\end{align}
$U$ and $U_R$ being the PMNS matrix, and another unitary matrix that diagonalize the active and sterile sectors respectively. %Hence, $U$ corresponds to the PMNS matrix. 
$\varphi$ is a diagonal matrix containing arbitrary phases $\varphi = {\rm diag}(e^{-i\phi_1},e^{-i\phi_2},e^{-i\phi_3})$, while $\mathbb{1}_3$ is the $3\times 3$ unitary matrix. A flavor neutrino field $\nu_{\beta L}$ ($\beta =e,\mu,\tau$) corresponds to a maximally-mixed superposition of two neutrino mass eigenstates $\nu_{k}^+$ and $\nu_{k}^-$, ($k=\{1,2,3\}$)~\cite{Kobayashi:2000md}
\begin{align}
    \nu_{\beta L}&=\frac{U_{\beta k}}{\sqrt{2}}(\nu_{k}^++i\,\nu_{k}^-)\, ,
\end{align}
having almost degenerate masses $m_{k, \pm}^2 = m_k^2 \pm \delta m_k^2/2$, respectively. For simplicity, we assume that mass difference $\delta m_k^2$, related to the matrix elements of $M_R$ and $Y$, is the same for all mass eigenstates, and simply write $\delta m^2$ hereafter. Current constraints indicate that $\delta m^2$ should be much smaller than the solar and atmospheric mass differences, $\delta m^2\ll |\Delta m_{21, 31}^2|$, and hence, over astrophysical baselines, oscillations induced by the former can happen whereas those due to the latter average out.
Thus, the flavor oscillation probability $P_{\beta\gamma}=P(\pbar{\nu_\beta}\to \pbar{\nu_\gamma})$  can be factorized in terms of an active-active survival probability $P_{aa}$ times the standard averaged term~\cite{deGouvea:2020eqq}
\begin{align}\label{eq:Prob}
    P_{\beta\gamma} = P_{aa}(E_\nu; L, \dm)\sum_{k} \left|U_{\beta k}\right|^2 \left|U_{\gamma k}\right|^2
\end{align}
where $E_\nu$ is the neutrino energy, and $L$ is the distance travelled. Neutrinos oscillations over astrophysical distances are also susceptible to decoherence due to separation of wave packets, owing to different group velocities of the mass-eigenstates. This is physically equivalent to an energy-dependent ``dephasing'' of the oscillation phase~\cite{Porto-Silva:2021ael}\footnote{We thank Georg Raffelt for pointing this out.}. Including such decoherence effects, $P_{aa}$ is
\begin{align}\label{eq:PDProb}
    P_{aa}(E_\nu) = \frac{1}{2}\left(1+e^{-\left(\frac{L}{L_{\rm coh}}\right)^2}\cos\left(\frac{2\pi L}{L_{\rm osc}}\right)\right).
\end{align}
The PD oscillation $L_{\rm osc}$ and coherence $L_{\rm coh}$ lengths have similar dependence on neutrino energy as in the standard case,
\begin{subequations}
\begin{align}
    L_{\rm osc} &= \frac{4\pi E_\nu}{\delta m^2}
    \approx 20\kpc\left(\frac{E_\nu}{25\MeV}\right)\left(\frac{10^{-19}\eV^2}{\delta m^2}\right),\\
    L_{\rm coh} &= \frac{4\sqrt{2} E_\nu}{|\delta m^2|}(E_\nu \sigma_x)\notag\\
    & \approx 114\kpc \left(\frac{E_\nu}{25\MeV}\right)^2\left(\frac{10^{-19}\eV^2}{\delta m^2}\right)\left(\frac{\sigma_x}{10^{-13}{\rm\ m}}\right),
\end{align}
\end{subequations}
where $\sigma_x$ is the initial size of the wave packet. We conclude that for $10^{-21}\eV^2~\lesssim \dm\lesssim 10^{-18}\eV^2$ the active-sterile oscillations can develop over scales of ${\cal O}({\rm kpc})$, right on the ballpark of expected baselines and energies for SN neutrinos.
The initial wave packet size can be determined from the processes producing the neutrinos in a SN, and has been estimated to be around $\sigma_x\sim 10^{-13}~{\rm m}$~\cite{Kersten:2015kio}, a value that we take as benchmark henceforth.
%%%%%%%%
%%%%%%%%%
\section{SN1987A Analysis and Results}
%%%%%%%%
%%%%%%%%%
The time-integrated neutrino spectra from a SN is well approximated by a blackbody emission, and is parameterized by the following alpha-fit spectra~\cite{Tamborra:2012ac}
\begin{align}\label{eq:Flux}
    \phi_\beta(E_\nu) = \frac{1}{E_{0\beta}}\frac{(1+\alpha)^{1+\alpha}}{\Gamma(1+\alpha)}\left(\frac{E_\nu}{E_{0\beta}}\right)^\alpha e^{-(1+\alpha)\frac{E_\nu}{E_{0\beta}}},
\end{align}
where $E_{0\beta}$ the average energy for a flavor $\nu_\beta$, and $\alpha$ is a parameter that determines the width of the distributions. For subsequent analysis, we set $\alpha=2.3$~\cite{Lunardini:2005jf}. We present in the Appendix for impact of having different values of $\alpha$.

Assuming that neutrinos are PD, the observed $\overline{\nu_e}$ fluence corresponds to the SN neutrino fluence multiplied by the PD probability, 
\begin{align}\label{eq:FluxEarth}
     \frac{d\Phi_{\rm 87}}{dE_\nu}= \frac{{\cal E}^{e}_{\rm tot}}{4\pi d^2}\,P_{aa}\left[\bar{p}\frac{\phi_{e}}{E_{0e}} + r_{xe}\,(1- \bar{p}) \frac{\phi_{x}}{E_{0x}}\right]\, \,.
\end{align}
where ${\cal E}^e_{\rm tot}$ is the SN total emitted energy in electron neutrinos (in ${\rm erg}$),  $r_{xe}={\cal E}^x_{\rm tot}/{\cal E}^e_{\rm tot}$ is a relative scale factor, and $d$ is the distance to Sanduleak -69 202, the progenitor star, taken here to be equal to 50 kpc. Here $\bar{p}=|U_{e1}|^2$ represents the permutation parameter between $\overline{\nu}_e$ and $\overline{\nu}_x$, $x$ being non-electron flavors, related to the adiabatic Mikheyev-Smirnov-Wolfenstein flavor conversions~\cite{PhysRevD.17.2369,Mikheev:1986gs,Dighe:1999bi}\footnote{Matter effects, both at the SN and the Earth, are not affected by the presence of the sterile states~\cite{Keranen:2004rg,Vissani:2014doa}}. 

\begin{figure}[h!]
	\centering
	\includegraphics[width=0.475\textwidth]{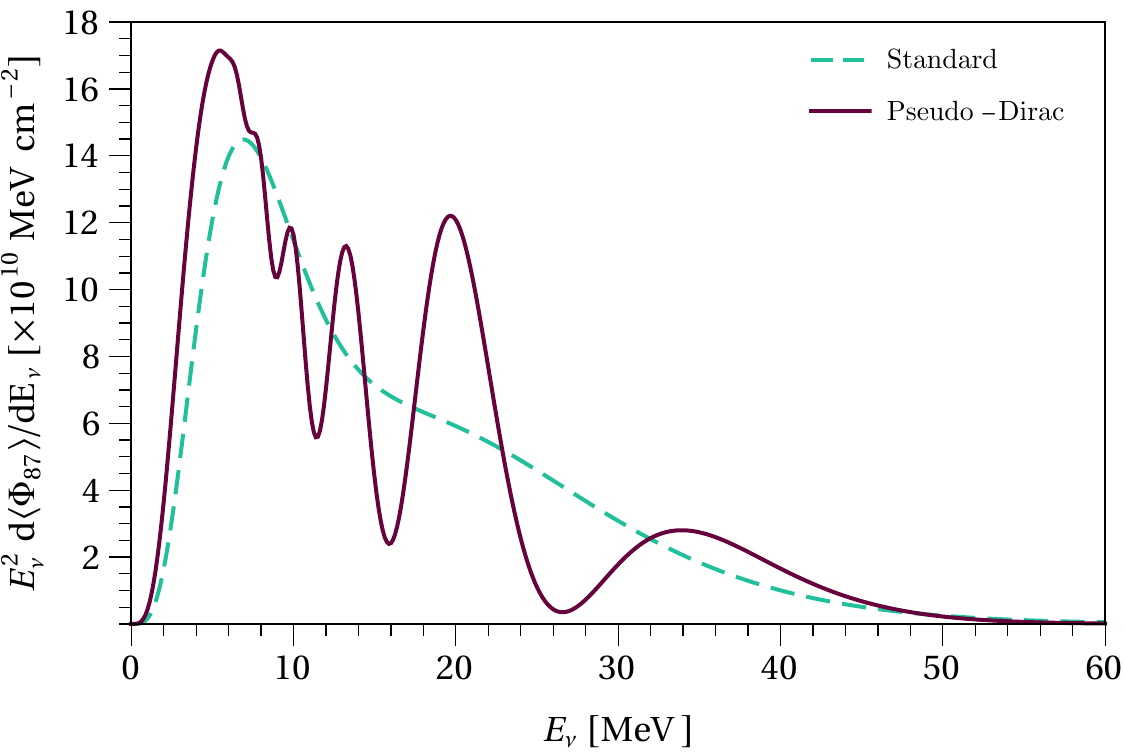}
	\caption{Energy averaged neutrino flux times energy squared, $E_\nu^2\, d\langle\Phi_{\rm 87}\rangle/dE_\nu$, as function of the neutrino energy for the standard case, i.e.~including only standard oscillations  (green dashed) and introducing active-sterile oscillations (purple). We assumed ${\cal E}^e_{\rm tot}=2.9\times 10^{53}~{\rm erg}$, $E_{0e}=4\MeV$, $E_{0x}=13\MeV$, and $\dm = 6.31\times 10^{-20}\eV^2$ in the PD case (purple), and ${\cal E}_{\rm tot}^e=1.2\times 10^{53}~{\rm erg}$, $E_{0e}=5\MeV$, $E_{0x}=15\MeV$  for the standard case (green).}
	\label{fig:E2Flux}
\end{figure}

In Fig.~\ref{fig:E2Flux}, we present the energy-averaged $\overline{\nu_e}$ flux multiplied by the energy squared, $E_\nu^2\, d\langle\Phi_{\rm 87}\rangle/dE_\nu$, for the standard case, i.e., including only standard oscillations  (green dashed) and introducing active-sterile oscillations, assuming $\dm = 6.31\times 10^{-20}\eV^2$ (purple). We have included a moderate neutrino energy resolution $\sigma_{E_\nu}=10\%/\sqrt{E_\nu/5 \MeV}$, only for illustration purposes. Interestingly, we observe that the oscillations induce a strong depletion in the flux for energies around 27.5 MeV. However, for $E_\nu\gtrsim 27.5\MeV$ the flux becomes larger than in the standard case.
%%%%%%%%%%%%%%%%%%%%%%%%%%%%%%%%%%%%%%%%%
%%%%%%%%%%%%%%%%%%%%%%%%%%%%%%%%%%%%%%%%%

To analyse the SN1987A data, we consider the events observed in KII, IMB and Baksan.
Since active-sterile oscillations mainly affect the neutrino energy spectra, we perform our analysis considering the fluence only. Following the standard treatments~\cite{Jegerlehner:1996kx,Mirizzi:2005tg,Lunardini:2005jf,Ianni:2009bd,Vissani:2014doa}, we define the unbinned extended likelihood for a single experiment as
\begin{align}
    \mathscr{L}&=e^{-N_{\rm tot}}\prod_{i}^{N_{\rm obs}}dE_i\left[\frac{dS}{dE_i}+\frac{dB}{dE_i}\right],
\end{align}
where $dS/dE_i\, (dB/dE_i)$ are the expected signal (background) events within an energy window $dE_i$ around the observed energy $E_i$, and $N_{\rm tot}$ ($N_{\rm obs}$) are the total number of expected (observed) events. We adopt the background treatment presented in \cite{Vissani:2014doa}, and we fix the minimum energy for KII equal to 4.5$\MeV$. In our analysis, we fit the fluence parameters $\{{\cal E}_{\rm tot}, E_{0e}, E_{0x}\}$ together with the quadratic mass difference $\delta m^2$, fixing the initial size of the neutrino wave packets to $\sigma_x=10^{-13}$~m. For the analysis, we fix $r_{xe}=1$, however, our results are not very sensitive to the variation in $r_{xe}$, as shown in the Appendix.

%%%%%%%%%%%%%%%%%%%%%%%%%%%%%%%%%%%%%%%%%
\begin{figure*}[t!]
\centering
\includegraphics[width=\textwidth]{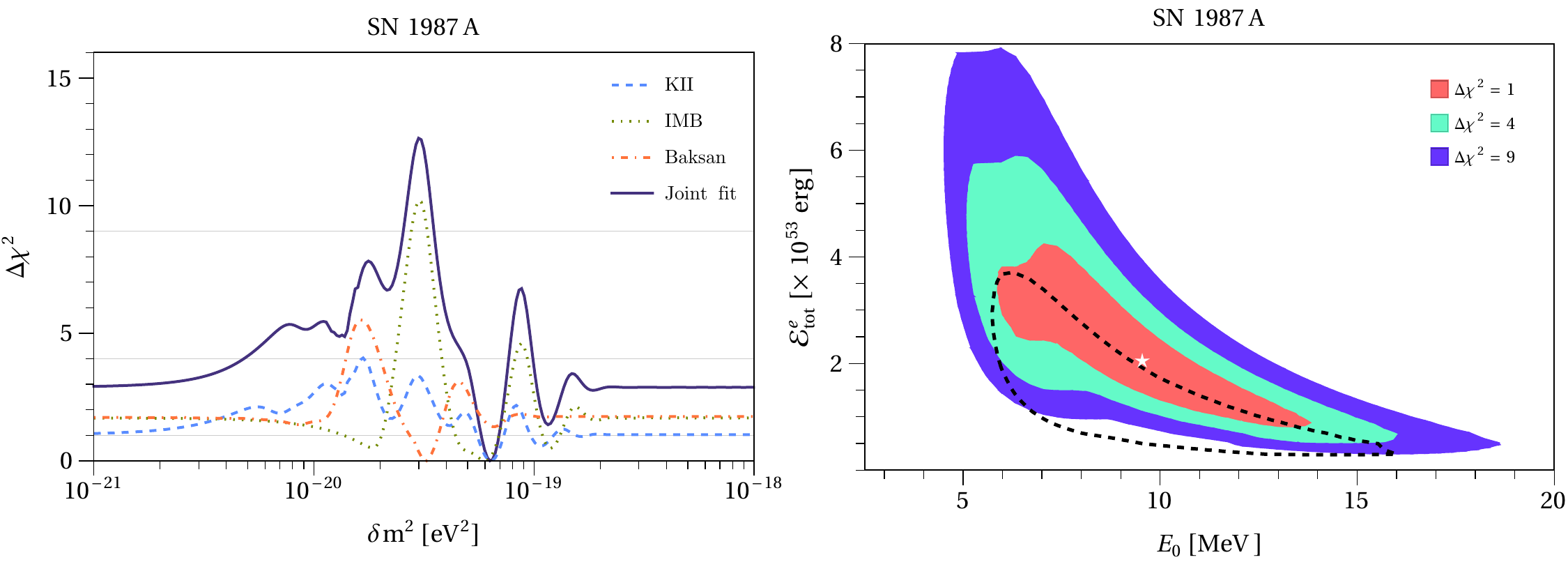}
\caption{Results of our analysis of the SN1987A data in the light of the PD scenario. Left: Marginalized $\Delta\chi^2$ as function of the quadratic mass difference $\dm$ for the individual analysis of KII (light-blue dashed), IMB (green dotted), Baksan (orange dot-dashed) and the combined analysis (purple). Right: Allowed regions for total energy ${\cal E}^e_{\rm tot}$ vs average energy, $E_0\equiv(E_{0e}+E_{0x})/2$. The black dashed is the allowed region without active-sterile oscillations at the $\Delta\chi^2 = 9$ level. \label{fig:chi287}}
\end{figure*}
%%%%%%%%%%%%%%%%%%%%%%%%%%%%%%%%%%%%%%%%%
%%%%%%%%%%%%%%%%%%%%%%%%%%%%%%%%%%%%%%%%%
\begin{table}
\caption{Best fit values of the fluence parameters, ${\cal E}^e_{\rm tot}$ in $10^{53}~{\rm erg}$, $E_{0e}$, $E_{0x}$ in $\MeV$,  and the quadratic mass difference $\dm$ in $10^{-20}~{\rm eV^2}$ for each individual experiment and their combination.
$\Delta\chi^2_{\rm No Osc}$ corresponds to the value at which the no-oscillation case is disfavored. \label{tab:BestFit}}
    \vspace{1mm}
    \centering
    \begin{tabular}{crrrrc}
        \toprule
        Experiment(s) & ${\cal E}^e_{\rm tot}$ & $E_{0e}$ &  $E_{0x}$ & $\dm$ & $\Delta\chi^2_{\rm No Osc}$\\ \colrule
        KII & $2.2$ & $4.24$ & $10.96$ & $6.31$ & 1.1\\ 
        IMB & $3.2$ & $1.36$ & $12.86$ & $6.03$ & 1.7\\ 
        Baksan & $15.7$ & $4.28$ & $8.03$ & $3.16$ & 1.7\\ 
        Joint Fit & $2.7$ & $4.00$ & $12.61$ & $6.31$ & 2.9\\\botrule
    \end{tabular}
\end{table}
%%%%%%%%%%%%%%%%%%%%%%%%%%%%%%%%%%%%%%%%%
In Fig.~\ref{fig:chi287} (left panel) we present $\Delta \chi^2\equiv -2(\ln{\mathscr{L}} - \ln{\mathscr{L}_{\rm max}})$ as function of $\dm$, marginalized over the fluence parameters, for each experiment, KII (light-blue dashed), IMB (green dotted), Baksan (orange dot-dashed), and the combined fit (purple), see also Tab.~\ref{tab:BestFit} for the best fit values in each case. Note that values of $\dm<10^{-21}\eV^2$ corresponds to the no-oscillation hypothesis, since for these values the oscillation length $\gtrsim 1\,{\rm Mpc}$, and is not relevant for a galactic core-collapse SN. Individually, KII prefers a non-zero $\dm$ with a $\Delta\chi^2\approx 1.1$ relative to the non-oscillated case, because the oscillated spectrum predicts less events in the energy window $E_i \sim 21\MeV - 31\MeV$, consistent with the data. Meanwhile, both IMB and Baksan have a larger preference for the PD scenario, with $\Delta\chi^2\approx 1.7$. However, the preferred values for both IMB and Baksan have a significant tension. Baksan prefers a value of $\dm\sim 3.3\times10^{-20}\eV^2$, such that its measured spectrum is enhanced around $E_i\sim 17\MeV$. Such value, nonetheless, predicts an oscillation minimum around $E_i\sim 35\MeV$, contrary to the IMB measurement.

%%%%%%%%%%%%%%%%%%%%%%%%%%%%%%%%%%%%%%%%%
\begin{figure}[t!]
\centering
\includegraphics[width=0.475\textwidth]{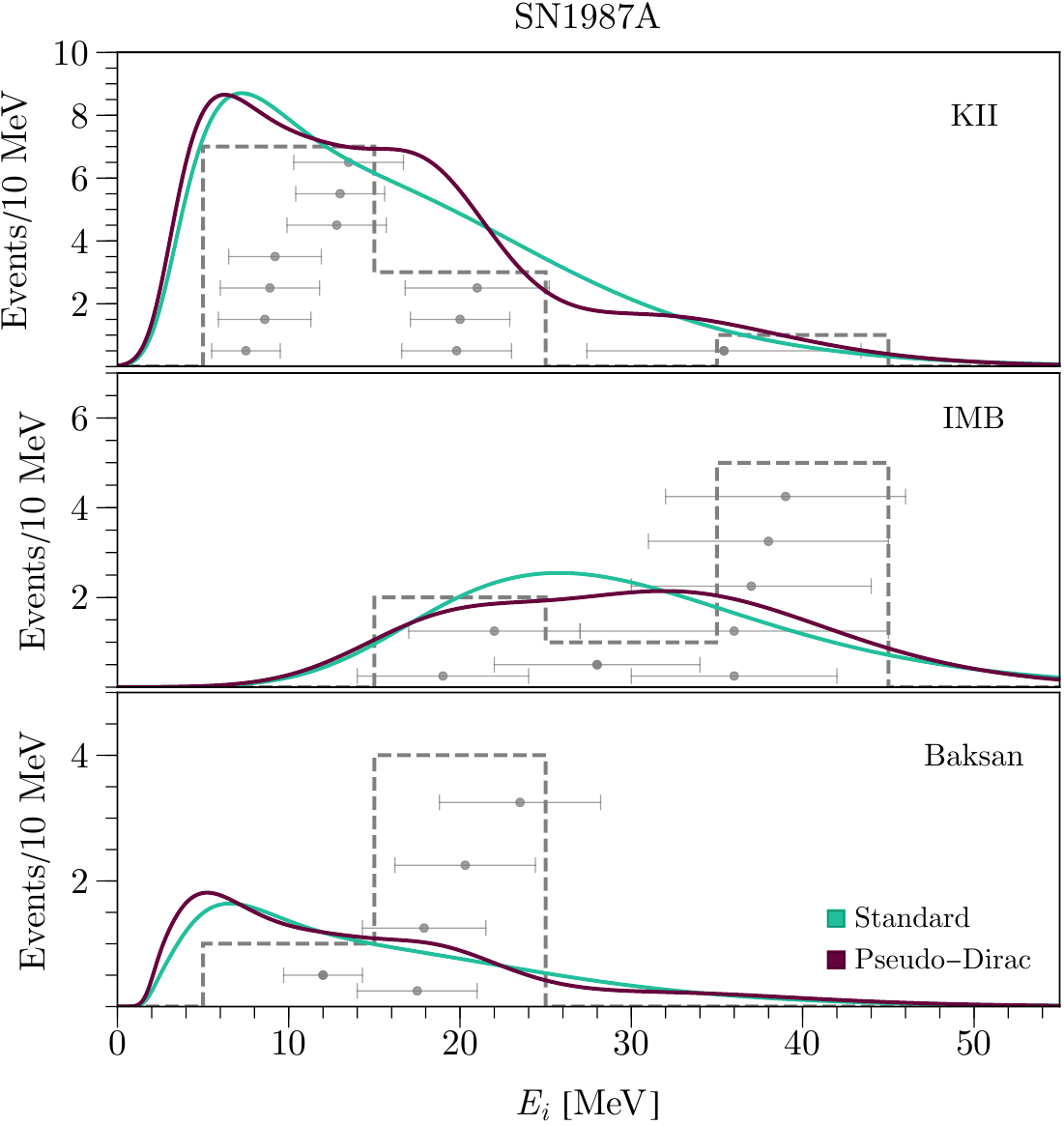}
\caption{Measured spectra (gray dashed lines) and individual observed events at KII (top), IMB (middle) and Baksan (bottom) as function of the positron energy $E_i$. We present the predicted spectra in the standard case (green) and including active-sterile oscillations (purple). The assumed parameters correspond to the best fit of the joint analysis, see Tab.~\ref{tab:BestFit} \label{fig:EvsSN87}}
\end{figure}
%%%%%%%%%%%%%%%%%%%%%%%%%%%%%%%%%%%%%%%%%
In the combined fit, we observe a preference for a non-zero value of $\dm = 6.31\times 10^{-20}\eV^2$, while the scenario without active-sterile oscillations ($\dm\lesssim 10^{-21}\eV^2$) is disfavored with $\Delta\chi^2\approx 3$. Such a preference is related to the slight tension between the events measured by KII and IMB: KII observed a spectra concentrated at lower energies, with an average positron energy of $\langle E_i \rangle_{\rm KII} = 15.4\pm 1.1\MeV$, than those measured at IMB ($\langle E_i \rangle_{\rm IMB} = 31.9\pm 2.3\MeV$), see Fig.~\ref{fig:EvsSN87}. Meanwhile, Baksan events are more compatible with KII ($\langle E_i \rangle_{\rm Baksan} = 18.2\pm 1.7\MeV$)\footnote{Although IMB has a small efficiency at lower energies, it observed more events than KII in the region where both experiments have similar efficiencies.}. In the absence of active-sterile oscillations, broad neutrino spectra are more favored to compensate this tension~\cite{Mirizzi:2005tg,Lunardini:2005jf}. The additional energy dependence coming from the oscillation probability $P_{aa}$ allows for a much broader spectrum in IMB, whilst still predicting for KII a reduction of events around $E_i\sim 27.5\MeV$, as seen in Fig.~\ref{fig:EvsSN87}.

Notably, the SN1987A data excludes values in the range $2.55\times 10^{-20}\eV^2\lesssim \dm \lesssim 3.0\times 10^{-20}\eV^2$, with $\Delta\chi^2 > 9$, the lowest quadratic mass values constrained by experiments so far. Clearly, a $\dm$ in such a region induces significant modifications to the spectra that contradict observations. For instance, for $\dm = 3.2\times 10^{-20}\eV^2$ --- excluded with $\Delta\chi^2 \approx 13$ ---, the KII spectrum would have more events in the window $E_i\in [15,25]\MeV$ than observed, while for IMB predicts almost no events with $E_i\gtrsim 35\MeV$, as noticed before. 

An interesting feature is present in the $\Delta\chi^2$ for values $\dm\gtrsim 4\times 10^{-19}\eV^2$: the obtained value of the $\Delta\chi^2$ is the same as in the case of no-oscillations. For such values of $\dm$, the active-sterile oscillations are averaged out, so that the detectable fluence arriving at the Earth is in fact half of the value originated at the SN1987A. Since in our simulation we have not included any prior, increasing the total energy by a factor of two can compensate the averaging out of the neutrino fluence. Thus, we obtain the same sensitivity as in the case without oscillations, consistent with some previous estimates~\cite{Midorikawa:1987kv,Midorikawa:1987rv,Esmaili:2012ac}. 

Finally, let us comment on the effects of the PD hypothesis on the observed fluence parameters from SN1987A. In Fig.~\ref{fig:chi287} (right panel), we present the allowed region in the total energy vs the average energy $E_0\equiv(E_{0e}+E_{0x})/2$ plane, marginalized with respect to the orthogonal parameter $\Delta E\equiv E_{0x}-E_{0e}$. The best fit prefers values of $E_0=9\MeV$, and somewhat larger values for ${\cal E}^e_{\rm tot}=2\times 10^{53}~{\rm erg}$. Although the PD scenario predicts a larger total energy, the regions are compatible with the non-oscillated case (black line) at the $\Delta\chi^2 \sim 9$ level.

\subsection{Future Sensitivity}
The next generation of neutrino experiments can probe the PD nature of neutrinos with a high accuracy. For our analysis, we consider two such detectors, DUNE~\cite{Abi:2020evt} and Hyper-Kamiokande (HK)~\cite{Abe:2018uyc}, owing to their large volume and the precision in the reconstruction of low energy neutrinos. The main interaction channel of MeV neutrinos at the DUNE liquid Argon detectors corresponds to $\nu_e$ scattering with Ar. To simulate the neutrino interaction in Liquid Argon detector, we used MARLEY, a Montecarlo event generator that allow to include all the nuclear transitions happening after the neutrino interaction~\cite{Gardiner:2021qfr}. On the other hand, HK is a water Cherenkov detector, which is mostly sensitive to $\overline{\nu}_{e}$ through the inverse beta decay process~\cite{Strumia:2003zx}. 
%Further details about our analysis are relegated to the Appendix.

We have considered that both the $\nu_{e}$ and the $\overline{\nu}_{e}$ components of the fluence follow an alpha fit (Eq.~\ref{eq:Flux}), and as benchmark scenario (the non-oscillation hypothesis), we have considered the best fit parameters describing SN1987A. In the analysis, the total energy (${\cal E}_{\rm tot}$), and the average energy ($E_{0}$) are treated as free parameters in the oscillation hypothesis, in that way we get a conservative prediction for the region that can be explored. The sensitivity plot is obtained by varying $\delta m^2$ for a given value of $\sigma_x$.

\begin{figure}[!t]
\includegraphics[width=.475\textwidth]{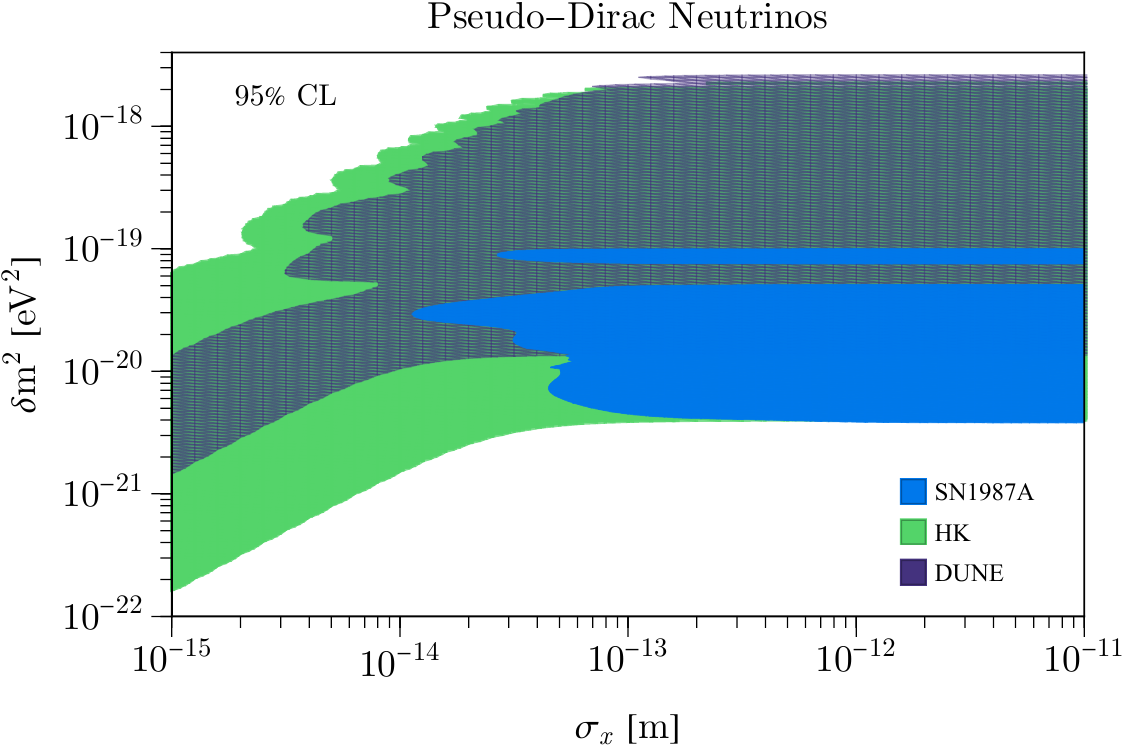}
\caption{Expected sensitivity on $\delta m^2$ as a function of $\sigma_{x}$ for DUNE (purple) and HK (green), together with the excluded region from SN1987A (blue) at more than $95\%$ CL. We assume that the SN occurs at 10 kpc.}
\label{fig:Future}
\end{figure}

Fig.~\ref{fig:Future} shows the $95\%$ CL sensitivity region for both DUNE and HK in the $\delta m^2-\sigma_x$ plane, for a SN happening at $10\kpc$, assuming a Gaussian likelihood analysis. For this baseline, the maximum sensitivity is expected to be around $\delta m^2 \sim 10^{-20}\eV^2$. For a smaller $\delta m^2$, neutrinos would not have enough time to oscillate by before arriving at the Earth. On the other hand, for a larger $\delta m^2$, wavepacket decoherence sets in. Although both experiments show a large overlap in the region that can be explored, there are some complementarities in their measurements. HK is sensitive to smaller values of $\delta m^2$ due to its larger volume, whereas DUNE can explore larger values of the mass splitting thanks to its energy resolution at the MeV scale. It is worth noting that both experiments can explore a large fraction of the excluded region from SN1987A. An interesting effect appears if the coherence length is smaller than the distance travelled by the neutrinos. That happens for $\sigma_{x} \leq 10^{-14}\text{m}$. In that case, the decoherence effect suppress the fluence at lower energies, as expected from Eq.~\eqref{eq:PDProb}, bringing an additional dependence on $\delta m^2$. 
%%%%%%%%%%%%%%%%%%%%%%%%%%%%%%%%%%%%%%%%%
%%%%%%%%%%%%%%%%%%%%%%%%%%%%%%%%%%%%%%%%%

%%%%%%%%
%%%%%%%%%
\section{Final Thoughts}
%%%%%%%%
%%%%%%%%%
To summarise, in this work, we explored the possibility that neutrinos are PD fermions, and the consequences of that on the neutrino fluence from a SN. We analyzed the neutrino data from SN1987A under such a hypothesis, and found a mild preference for PD nature of neutrinos, owing to the slight tension between the data from KII and IMB. Interestingly, mass-squared differences between $2.55\times 10^{-20}\eV^2\lesssim \dm \lesssim 3.0\times 10^{-20}\eV^2$ are excluded with $\Delta\chi^2 > 9$, resulting on the first constraint yet on such tiny mass differences. The next galactic SN will be a watershed moment in the history of neutrino physics. Equipped with the next generation detectors, one would expect to detect tens of thousands of events, which can be leveraged to put strong bounds on extreme neutrino properties. Our study reveals that for a future galactic SN happening at $10\kpc$, DUNE and HK can easily probe tiny mass-squared differences, $\dm \sim 10^{-20}\eV^2$. One may wonder, if it is possible to falsify such a scenario. We believe that neutral current measurements will play a crucial role in testing this scenario, since these active-sterile oscillations would induce a disappearance of all flavor states in the same way. Observations of non-electron neutrino events (see, e.g.,~\cite{Lang:2016zhv}) in future detectors can definitely shed more light on this topic.    

%%%%%%%%%%%%%%%%%%%%%%%%%%%%%%%%%%%%%%%%%
\acknowledgments
%%%%%%%%%%%%%%%%%%%%%%%%%%%%%%%%%%%%%%%%%

%\emph{Acknowledgments.}--- 
We are grateful to Basudeb Dasgupta, Andr\'e de Gouv\^ea, Cecilia Lunardini, Pedro Machado, and Georg Raffelt, for a careful reading of the manuscript, and a number of insightful comments. We acknowledge useful discussions with Steven Gardiner regarding MARLEY. This manuscript has been authored by Fermi Research Alliance, LLC under Contract No. DE-AC02-07CH11359 with the U.S. Department of Energy, Office of Science, Office of High Energy Physics. MS acknowledges support from the National Science Foundation, Grant PHY-1630782, and to the Heising-Simons Foundation, Grant 2017-228.

\appendix

%In this Supplemental Material, we provide additional details, which may be relevant to experts. Specifically, we discuss the dependence of our results on the pinching parameter and the initial wave packet size. Moreover, we provide additional details regarding the future sensitivity simulations.

%%%%%%%%%%%%%%%%%%%%%%%%%%%
\section{Additional Cross-checks}

\subsection{Dependence on $\langle E_{0e,0x} \rangle$}

\begin{figure}[h!]
\centering
\includegraphics[width=0.48\textwidth]{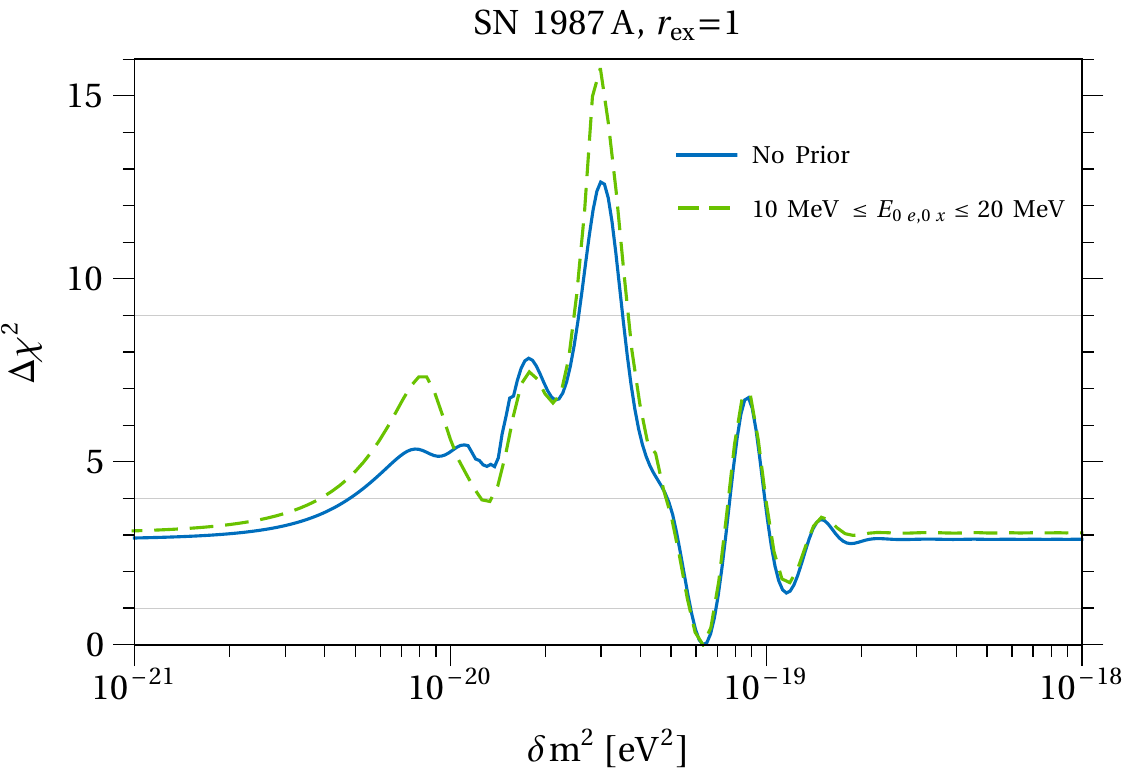}
\caption{Marginalized $\Delta\chi^2$ for the joint fit of the three distinct data sets coming from KII, IMB and Baksan as function of $\dm$ for different priors on $\langle E_{0e,0x}\rangle$. \label{fig:chiEex}}
\end{figure}

Supernova simulations have established that the average neutrino energies should be of $\mathcal{O}(10)$~MeV, while the standard SN1987a prefers much smaller values for $E_{0e}$ (see Tab. I). Thus, one may wonder if the results presented in the main text would suffer any alteration if we impose a prior on the average energies of the $\nu_{e,x}$. In Fig.~\ref{fig:chiEex}, we present the results of our analysis after introducing a flat prior on $E_{0e,\, 0x} \in [10,20]$~MeV. We observe no significant modification on the $\Delta\chi^2$, apart from the region $\dm\in[5,30]\times 10^{-21}\eV^2$. Our best fit in this case corresponds to $\dm=6.31\times10^{-20}\eV^2$, ${\cal E}_{\rm tot}=0.54\times 10^{53}$~erg, $E_{0e}=10\MeV$, and $E_{0x}=15.05\MeV$. 
%%%%%%%%%%%%%%%%%%%%%%%%%%%%%%%%%%%%%%%%%
%%%%%%%%%%%%%%%%%%%%%%%%%%%%%%%%%%%%%%%%%

\subsection{Dependence on $r_{xe}={\cal E}^x_{\rm tot}/{\cal E}^e_{\rm tot}$}
\begin{figure}[h!]
\centering
\includegraphics[width=0.48\textwidth]{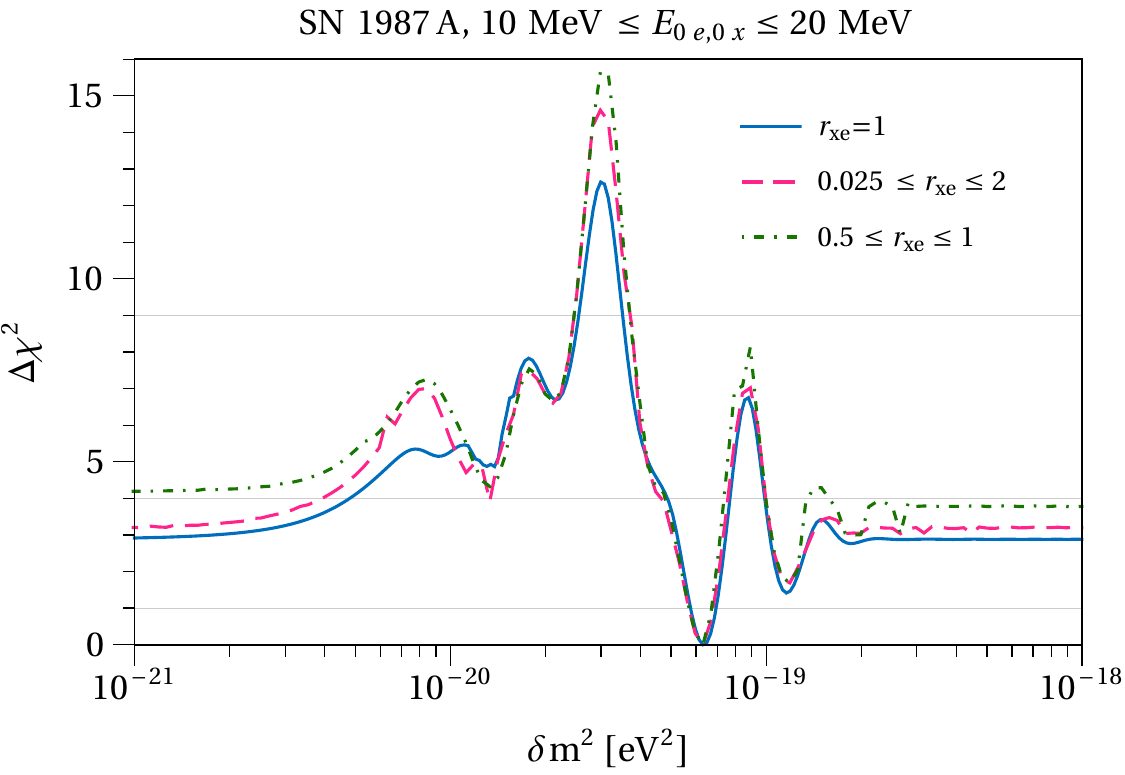}
\caption{Marginalized $\Delta\chi^2$ for the joint fit of the three distinct data sets coming from KII, IMB and Baksan as function of $\dm$ for different values of $r_{xe}$ as discussed in the text. \label{fig:chirx}}
\end{figure}

%%%%%%%%%%%%%%%%%%%%%%%%%%%%%%%%%%%%%%%%%
\begin{table}
\caption{Best fit values of the fluence parameters, ${\cal E}^e_{\rm tot}$ in $10^{53}~{\rm erg}$, $E_{0e}$, $E_{0x}$ in $\MeV$, the quadratic mass difference $\dm$ in $10^{-20}~{\rm eV^2}$, and the $\nu_x$ to $\nu_e$ total energy ratio $r_{ex}$, for two distinct priors. \label{tab:BF-check}}
    \vspace{1mm}
    \centering
    \begin{tabular}{crrrrc}
        \toprule
        Prior & ${\cal E}^e_{\rm tot}$ & $E_{0e}$ &  $E_{0x}$ & $\dm$ & $r_{ex}$\\ \colrule
        B & $0.68$ & $10.00$ & $16.96$ & $6.31$ & $0.48$\\ 
        C & $0.58$ & $11.52$ & $11.53$ & $6.31$ & $0.99$\\\botrule
    \end{tabular}
\end{table}
%%%%%%%%%%%%%%%%%%%%%%%%%%%%%%%%%%%%%%%%%
In this subsection, we discuss the sensitivity of our analysis to changes in values of $r_{xe}={\cal E}^x_{\rm tot}/{\cal E}^e_{\rm tot}$. In Fig.\,\ref{fig:chirx}, we show the combined $\Delta\chi^2$ for three different cases, (A) $r_{xe}=1$, (B) $0.025\leq r_{xe}\leq 2$, and (C) $0.5\leq r_{xe}\leq 1$
The first case considers $r_{xe}$ to be fixed, while in the latter two cases, we marginalize over $r_{xe}$.
We present the best fit in each case in Table~\ref{tab:BF-check} for the combined analysis.
We find that our results are mildly dependent on the variation in $r_{xe}$, and does not affect our best fit point. In fact, for the third case, we find that the no-oscillation hypothesis is ruled out with a slightly larger $\Delta \chi^2\gtrsim 4$. Since the dependence in very mild, we present our results in terms of $r_{xe}=1$ in our main text.
%%%%%%%%%%%%%%%%%%%%%%%%%%%%%%%%%%%%%%%%%
%%%%%%%%%%%%%%%%%%%%%%%%%%%%%%%%%%%%%%%%%

\subsection{Dependence on the pinching parameter}

\begin{figure}[h!]
\centering
\includegraphics[width=0.48\textwidth]{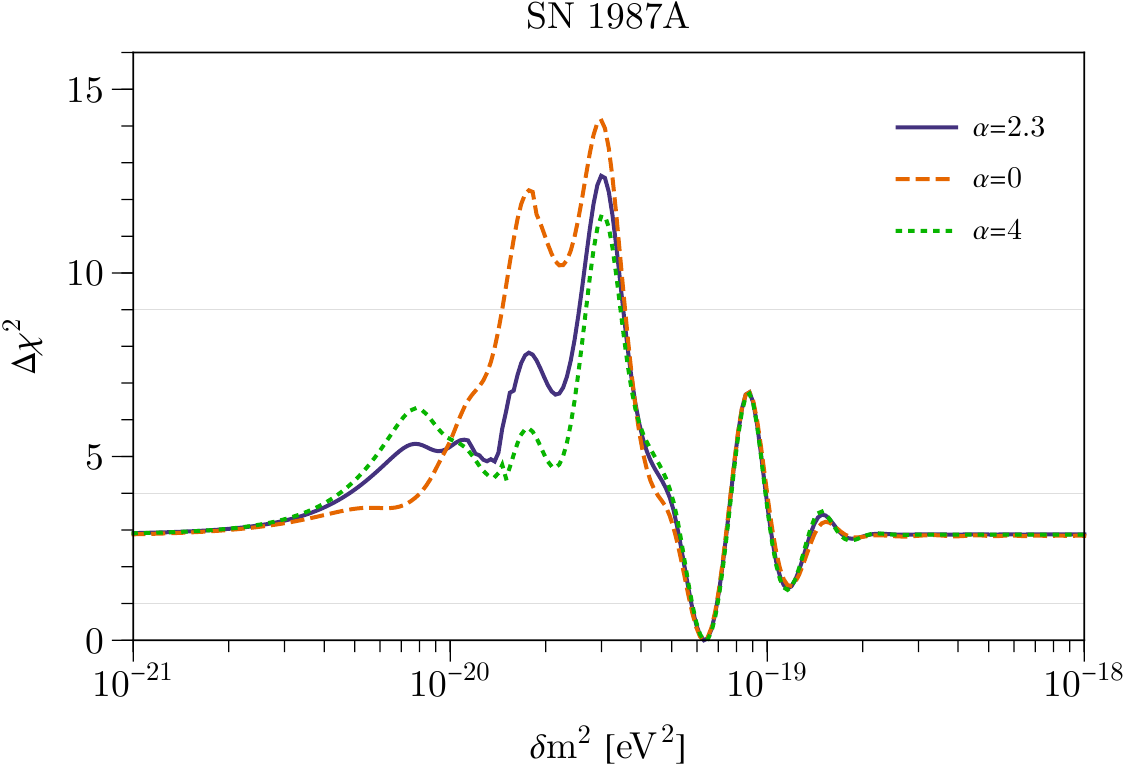}
\caption{Marginalized $\Delta\chi^2$ for the joint fit of the three distinct data sets coming from KII, IMB and Baksan as function of $\dm$ for different values of the pinching parameter $\alpha=$ 0 (orange dashed), 2.3 (purple), and 4 (green dotted). \label{fig:chialp}}
\end{figure}

The pinching parameter $\alpha$ is a crucial parameter which describes how the SN neutrino spectrum departs from being fully thermal. In our simulation, we have fixed as prior a value of $\alpha =2.3$, resulting from different SN simulations. Here we present how our results change if we choose a different value instead. In Fig.~\ref{fig:chialp}, we present the combined $\Delta\chi^2$ for three different values of $\alpha = \{0., 2.3, 4\}$. We observe that our results are only mildly dependent on the pinching parameter. In fact, for values of the mass-squared difference $\dm\gtrsim 4\times 10^{-20}\eV^2$, the marginalized fit basically coincides for all values. Moreover, the preference for $\dm = 6.31\times 10^{-20}\eV^2$ is independent of $\alpha$. For values $\dm\lesssim 4\times 10^{-20}\eV^2$, larger differences are present, but the overall behavior is the same.

\subsection{Varying the initial wave packet size}

\begin{figure}[h!]
\centering
\includegraphics[width=0.48\textwidth]{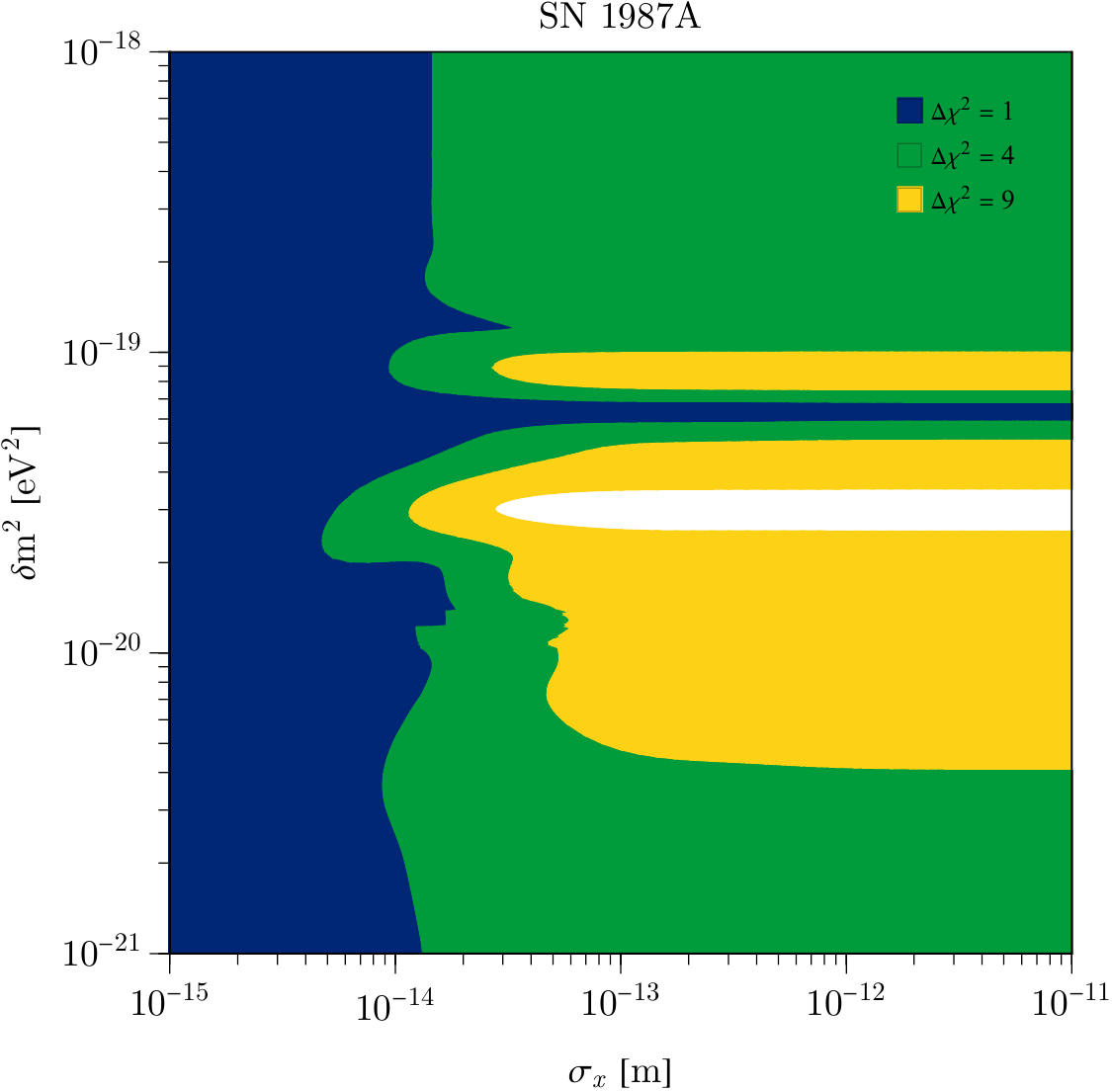}
\caption{Marginalized $\Delta\chi^2$ for the joint fit of the three distinct data sets coming from KII, IMB and Baksan in the plane $\dm$ vs $\sigma_x$. The regions with $\Delta\chi^2 \le \{1,4,9\}$ correspond to the blue, green, and yellow colors respectively. The white region is excluded at more than $\Delta\chi^2 >9.$ \label{fig:chisx}}
\end{figure}

Another important parameter corresponds to the initial wave packet size $\sigma_x$. As pointed out in the main text, estimates indicate that such values should be around $\sigma_x = 10^{-13}~{\rm m}$. Nevertheless, there is no definite consensus on what is its actual value. Here, we consider how our results change by modifying the value of $\sigma_x$. Let us stress that we are \emph{not} varying $\sigma_x$ as a parameter to be fitted; instead we fix $\sigma_x$ and we determine the $\Delta\chi^2$ in each case. In Fig.~\ref{fig:chisx}, we present our results. We observe that for values $\sigma_x\gtrsim 2\times 10^{-13}{\rm~m}$, the $\Delta\chi^2$ is independent of the initial wave packet size. Such independence arises because the coherence lengths become much larger than the distance traveled by the neutrinos. In other words, the neutrino wave packets do not have decoherence in this case. On the other hand, when the decoherence is important, that is, for $\sigma_x\lesssim 2\times 10^{-14}{\rm~m}$, the oscillations are erased, and the sensitivity to active-sterile oscillation is lost. Explicitly, for $\sigma_x\lesssim 4\times 10^{-15}{\rm~m}$, the $\Delta\chi^2 \le 1$ for all values of $\dm$ in the range that we are considering. Since for such values the decoherence lengths are smaller than the oscillations lengths, the flux that arrives to the Earth is basically an incoherent superposition, so that the active component of the flux is close to half the original value. Therefore, given that in the fit the reduction of the flux can be compensated by increasing the total energy, the sensitivity to the PD scenario is lost.

\subsection{Future Sensitivity Details}

As shown in previous sections, in the case of a core-collapse
supernova, the next generation of experiments, will probe mass splittings between active and sterile along with several orders of magnitude. To observe neutrinos at the MeV scale, in the future, there would be two ideal detectors, DUNE~\cite{Abi:2020evt} and Hyper-Kamiokande (HK)~\cite{Abe:2018uyc}, thanks to their large volume
and the precision in the reconstruction of low energy neutrinos. As a benchmark scenario, we have considered a supernova happening
at 10~kpc described by the best-fit parameters of the SN1987A.

\begin{figure*}[t!]
\centering
\includegraphics[width=0.48\textwidth]{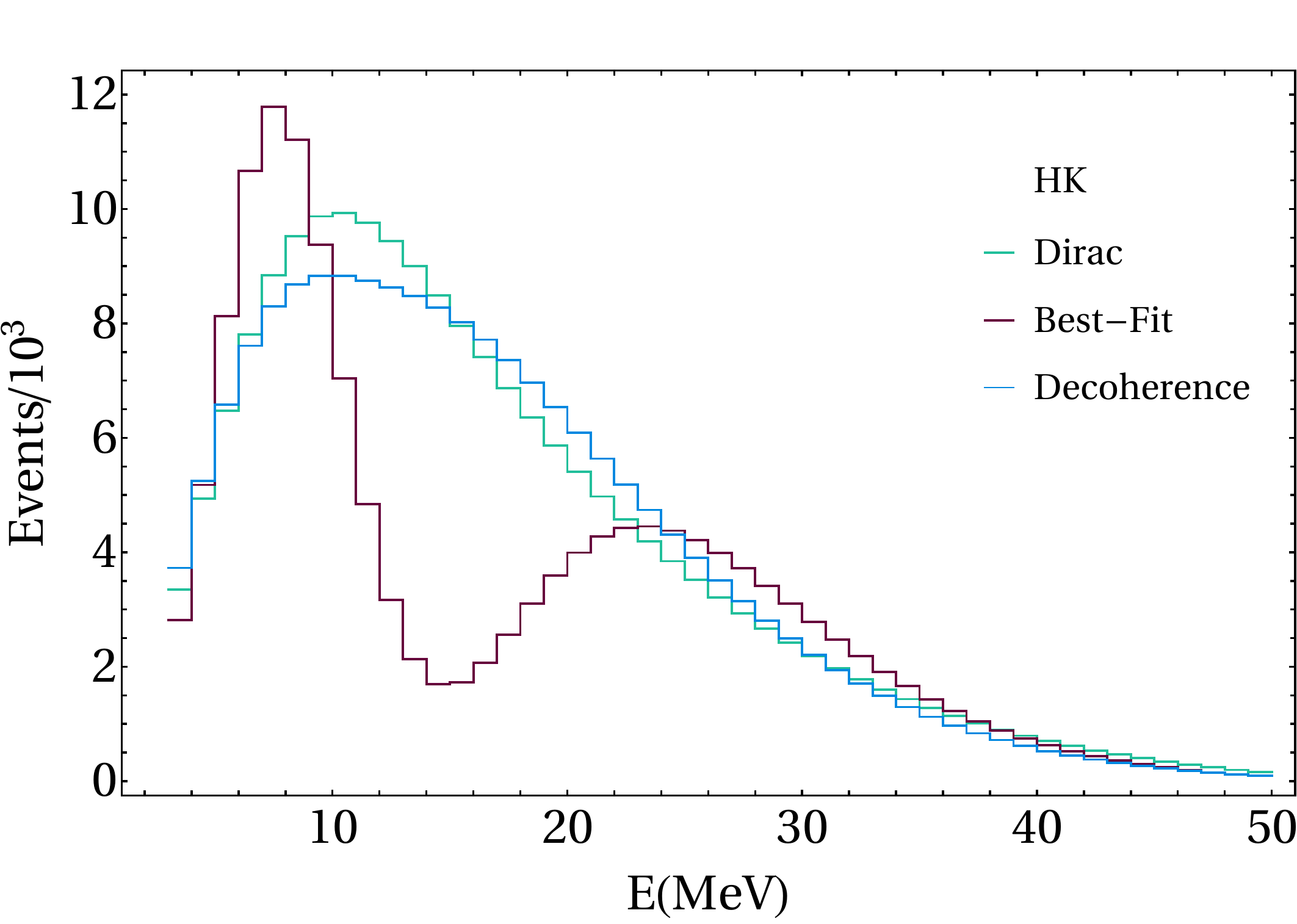}
\includegraphics[width=0.48\textwidth]{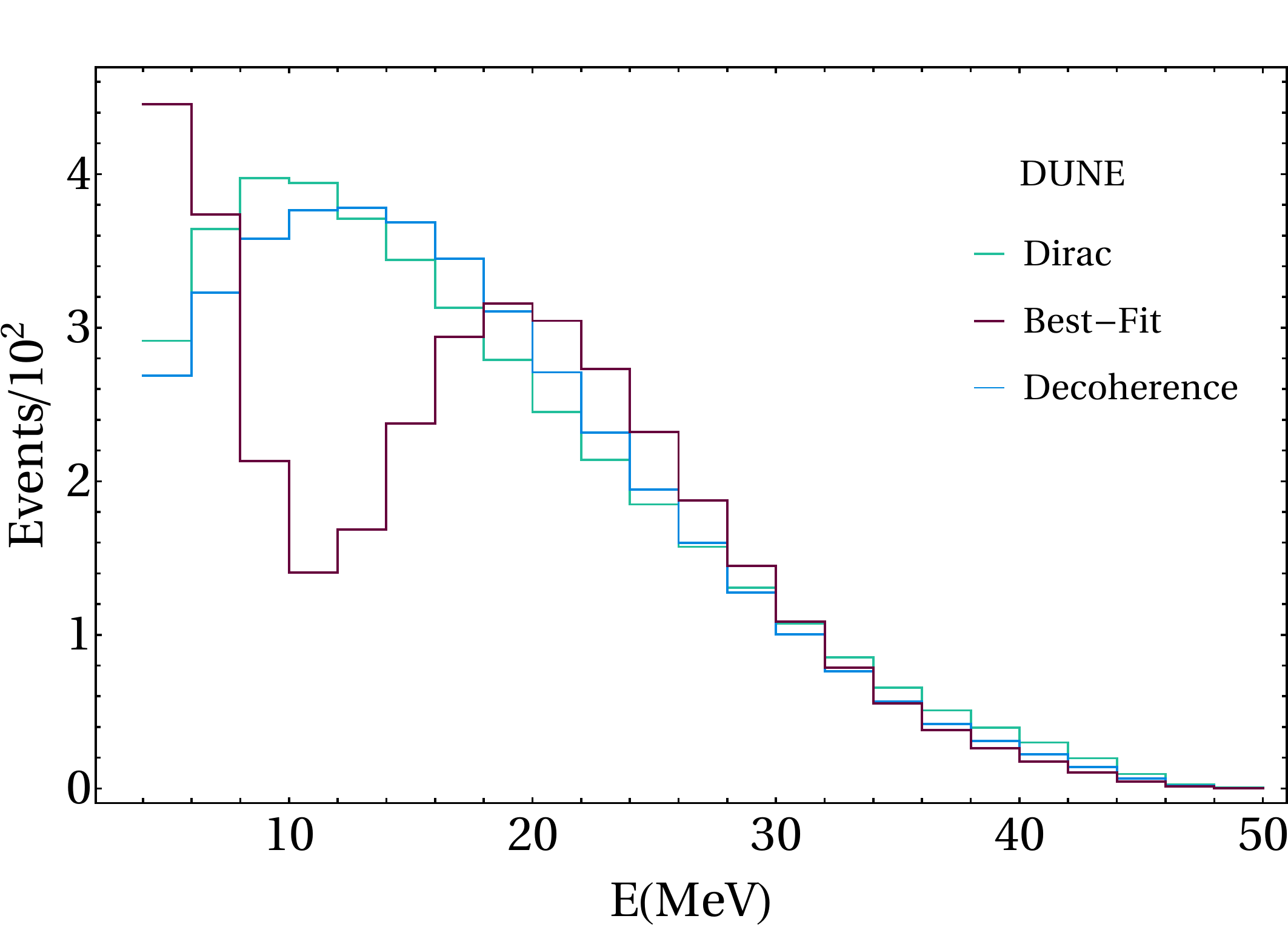}
\caption{The Number of events expected in HK (left) and DUNE (right) for a supernova happening at 10kpc. For the supernova luminosity, we assume the best-fit value of the SN1987A. We show the number of events for three different scenarios: neutrinos are Dirac fermions, the best-fit point of the SN1987A analysis, and coherence lengths shorter than 10kpc (Decoherence). In particular, in the last case, we use the following parameters: $\delta m^2 = 5\times10^{-21}\text{eV}^2, \sigma_{x} = 10^{-15}$~m \label{fig:FEvnt}}
\end{figure*}

In a Liquid Argon detector, the main interaction channel of MeV
neutrinos consists of the scattering of electron neutrinos with the
Argon nuclei ($\nu_e + ^{40}{\rm Ar}\rightarrow e^{-} + ^{40}{\rm K}^{\ast}$),
which will generate an electron and an excited nucleus of potassium
($^{40}{\rm K}^{\ast}$). After the de-excitation of the potassium, a photon cascade is also generated. Other particles can also be generated in the neutrino interaction, like neutrons, protons, or deuterons. In this work, we will consider just the observation of the electron. To simulate the neutrino interaction in the Liquid Argon detector, we used MARLEY, a Montecarlo event generator, to include all the nuclear transitions happening after the neutrino interaction~\cite{Gardiner:2021qfr}. After a neutrino interaction, we consider that it can be detected if an electron with energy larger than 4~MeV is generated. The energy associated to each event correspond to the reconstructed electron energy, and they will be distributed in energy bins of 2~MeV. The finite energy resolution of the detector will introduce an error in the reconstruction of the electron energy. Following the
measurement done in previous Liquid Argon experiments~\cite{Amoruso:2003sw}, we assumed an increase in the
resolution energy as $\sigma_{E} = 0.11\sqrt{E/\text{MeV}} + 0.2
(E/\text{MeV})$. As a fiducial volume, we assumed 40~ktons for DUNE. The expected number of events is shown in Fig.\ref{fig:FEvnt} for three different cases: neutrinos are pure Dirac states (Dirac), the best-fit of the SN1987A analysis ($\delta m^2=6.31\times 10^{-20}$ and $\sigma_x = 10^{-13}$~m), and the coherence length is shorter than the distance traveled by the neutrinos (Decoherence). In the last case, we have set $\delta m^2=5\times 10^{-21}\text{eV}^2$ and $\sigma_{x}=10^{-15}$~m. If neutrinos are pseudo-Dirac particles, and the mass splitting is on order $\sim 10^{-20}\text{eV}^2$ an oscillation pattern would be observed in the event distribution. In the case where the size of the wave packet is small enough such that neutrinos will arrive as an incoherent superposition of states, there would still be a sensitivity over the mass splitting due to the modifications of the flux at lower energies. The expected sensitivity for DUNE for different values of $\sigma_x$ is shown in Fig. 4 of the main paper.

A water Cherenkov detector will mainly observe the electron-antineutrino component of the supernova. At the MeV scale, $\overline{\nu}_{e}$ interact via Inverse Beta Decay (IBD)~\cite{Strumia:2003zx} with the free protons in water ($\overline{\nu}_{e}+p\rightarrow e^{+}+n$). In the reconstruction of the positron energy, we have assumed a similar energy resolution as Super-Kamiokande~\cite{Abe:2018uyc} ($\sigma_{E}=0.6\sqrt{E/\text{MeV}}$) in the measurement of solar neutrinos. The uncertainty in the energy is included assuming a Gaussian distribution of the energy measured centered at the electron energy after the interaction. The events are distributed as a function of the reconstructed positron energy in bins of 1~MeV, Fig.~\ref{fig:FEvnt} (left). The deviations found in the number of events is similar to the DUNE experiment. As a fiducial volume for HK, we have considered one tank of 187~kton. The sensitivity of HK for different values of $\sigma_x$ is shown in Fig. 4 of the main paper.

% \bibliographystyle{apsrev4-1}
% \bibliography{PDSN}{}

%\end{document}

%\bibliographystyle{kpmod}
\bibliographystyle{apsrev4-1}
\bibliography{PDSN.bib}

\begin{thebibliography}{61}%
\makeatletter
\providecommand \@ifxundefined [1]{%
 \@ifx{#1\undefined}
}%
\providecommand \@ifnum [1]{%
 \ifnum #1\expandafter \@firstoftwo
 \else \expandafter \@secondoftwo
 \fi
}%
\providecommand \@ifx [1]{%
 \ifx #1\expandafter \@firstoftwo
 \else \expandafter \@secondoftwo
 \fi
}%
\providecommand \natexlab [1]{#1}%
\providecommand \enquote  [1]{``#1''}%
\providecommand \bibnamefont  [1]{#1}%
\providecommand \bibfnamefont [1]{#1}%
\providecommand \citenamefont [1]{#1}%
\providecommand \href@noop [0]{\@secondoftwo}%
\providecommand \href [0]{\begingroup \@sanitize@url \@href}%
\providecommand \@href[1]{\@@startlink{#1}\@@href}%
\providecommand \@@href[1]{\endgroup#1\@@endlink}%
\providecommand \@sanitize@url [0]{\catcode `\\12\catcode `\$12\catcode
  `\&12\catcode `\#12\catcode `\^12\catcode `\_12\catcode `\%12\relax}%
\providecommand \@@startlink[1]{}%
\providecommand \@@endlink[0]{}%
\providecommand \url  [0]{\begingroup\@sanitize@url \@url }%
\providecommand \@url [1]{\endgroup\@href {#1}{\urlprefix }}%
\providecommand \urlprefix  [0]{URL }%
\providecommand \Eprint [0]{\href }%
\providecommand \doibase [0]{http://dx.doi.org/}%
\providecommand \selectlanguage [0]{\@gobble}%
\providecommand \bibinfo  [0]{\@secondoftwo}%
\providecommand \bibfield  [0]{\@secondoftwo}%
\providecommand \translation [1]{[#1]}%
\providecommand \BibitemOpen [0]{}%
\providecommand \bibitemStop [0]{}%
\providecommand \bibitemNoStop [0]{.\EOS\space}%
\providecommand \EOS [0]{\spacefactor3000\relax}%
\providecommand \BibitemShut  [1]{\csname bibitem#1\endcsname}%
\let\auto@bib@innerbib\@empty
%</preamble>
\bibitem [{\citenamefont {Kayser}\ and\ \citenamefont
  {Shrock}(1982)}]{Kayser:1981nw}%
  \BibitemOpen
  \bibfield  {author} {\bibinfo {author} {\bibfnamefont {B.}~\bibnamefont
  {Kayser}}\ and\ \bibinfo {author} {\bibfnamefont {R.~E.}\ \bibnamefont
  {Shrock}},\ }\href {\doibase 10.1016/0370-2693(82)90314-8} {\bibfield
  {journal} {\bibinfo  {journal} {Phys. Lett. B}\ }\textbf {\bibinfo {volume}
  {112}},\ \bibinfo {pages} {137} (\bibinfo {year} {1982})}\BibitemShut
  {NoStop}%
\bibitem [{\citenamefont {Furry}(1938)}]{Furry:1938zz}%
  \BibitemOpen
  \bibfield  {author} {\bibinfo {author} {\bibfnamefont {W.~H.}\ \bibnamefont
  {Furry}},\ }\href {\doibase 10.1103/PhysRev.54.56} {\bibfield  {journal}
  {\bibinfo  {journal} {Phys. Rev.}\ }\textbf {\bibinfo {volume} {54}},\
  \bibinfo {pages} {56} (\bibinfo {year} {1938})}\BibitemShut {NoStop}%
\bibitem [{\citenamefont {Long}\ \emph {et~al.}(2014)\citenamefont {Long},
  \citenamefont {Lunardini},\ and\ \citenamefont {Sabancilar}}]{Long:2014zva}%
  \BibitemOpen
  \bibfield  {author} {\bibinfo {author} {\bibfnamefont {A.~J.}\ \bibnamefont
  {Long}}, \bibinfo {author} {\bibfnamefont {C.}~\bibnamefont {Lunardini}}, \
  and\ \bibinfo {author} {\bibfnamefont {E.}~\bibnamefont {Sabancilar}},\
  }\href {\doibase 10.1088/1475-7516/2014/08/038} {\bibfield  {journal}
  {\bibinfo  {journal} {JCAP}\ }\textbf {\bibinfo {volume} {08}},\ \bibinfo
  {pages} {038} (\bibinfo {year} {2014})},\ \Eprint
  {http://arxiv.org/abs/1405.7654} {arXiv:1405.7654 [hep-ph]} \BibitemShut
  {NoStop}%
\bibitem [{\citenamefont {Berryman}\ \emph {et~al.}(2018)\citenamefont
  {Berryman}, \citenamefont {de~Gouv\^ea}, \citenamefont {Kelly},\ and\
  \citenamefont {Schmitt}}]{Berryman:2018qxn}%
  \BibitemOpen
  \bibfield  {author} {\bibinfo {author} {\bibfnamefont {J.~M.}\ \bibnamefont
  {Berryman}}, \bibinfo {author} {\bibfnamefont {A.}~\bibnamefont
  {de~Gouv\^ea}}, \bibinfo {author} {\bibfnamefont {K.~J.}\ \bibnamefont
  {Kelly}}, \ and\ \bibinfo {author} {\bibfnamefont {M.}~\bibnamefont
  {Schmitt}},\ }\href {\doibase 10.1103/PhysRevD.98.016009} {\bibfield
  {journal} {\bibinfo  {journal} {Phys. Rev. D}\ }\textbf {\bibinfo {volume}
  {98}},\ \bibinfo {pages} {016009} (\bibinfo {year} {2018})},\ \Eprint
  {http://arxiv.org/abs/1805.10294} {arXiv:1805.10294 [hep-ph]} \BibitemShut
  {NoStop}%
\bibitem [{\citenamefont {Millar}\ \emph {et~al.}(2018)\citenamefont {Millar},
  \citenamefont {Raffelt}, \citenamefont {Stodolsky},\ and\ \citenamefont
  {Vitagliano}}]{Millar:2018hkv}%
  \BibitemOpen
  \bibfield  {author} {\bibinfo {author} {\bibfnamefont {A.}~\bibnamefont
  {Millar}}, \bibinfo {author} {\bibfnamefont {G.}~\bibnamefont {Raffelt}},
  \bibinfo {author} {\bibfnamefont {L.}~\bibnamefont {Stodolsky}}, \ and\
  \bibinfo {author} {\bibfnamefont {E.}~\bibnamefont {Vitagliano}},\ }\href
  {\doibase 10.1103/PhysRevD.98.123006} {\bibfield  {journal} {\bibinfo
  {journal} {Phys. Rev. D}\ }\textbf {\bibinfo {volume} {98}},\ \bibinfo
  {pages} {123006} (\bibinfo {year} {2018})},\ \Eprint
  {http://arxiv.org/abs/1810.06584} {arXiv:1810.06584 [hep-ph]} \BibitemShut
  {NoStop}%
\bibitem [{\citenamefont {Balantekin}\ \emph {et~al.}(2019)\citenamefont
  {Balantekin}, \citenamefont {de~Gouv\^ea},\ and\ \citenamefont
  {Kayser}}]{Balantekin:2018ukw}%
  \BibitemOpen
  \bibfield  {author} {\bibinfo {author} {\bibfnamefont {A.~B.}\ \bibnamefont
  {Balantekin}}, \bibinfo {author} {\bibfnamefont {A.}~\bibnamefont
  {de~Gouv\^ea}}, \ and\ \bibinfo {author} {\bibfnamefont {B.}~\bibnamefont
  {Kayser}},\ }\href {\doibase 10.1016/j.physletb.2018.11.068} {\bibfield
  {journal} {\bibinfo  {journal} {Phys. Lett.}\ }\textbf {\bibinfo {volume}
  {B789}},\ \bibinfo {pages} {488} (\bibinfo {year} {2019})},\ \Eprint
  {http://arxiv.org/abs/1808.10518} {arXiv:1808.10518 [hep-ph]} \BibitemShut
  {NoStop}%
%%CITATION = ARXIV:1808.10518;%%
\bibitem [{\citenamefont {Funcke}\ \emph {et~al.}(2020)\citenamefont {Funcke},
  \citenamefont {Raffelt},\ and\ \citenamefont {Vitagliano}}]{Funcke:2019grs}%
  \BibitemOpen
  \bibfield  {author} {\bibinfo {author} {\bibfnamefont {L.}~\bibnamefont
  {Funcke}}, \bibinfo {author} {\bibfnamefont {G.}~\bibnamefont {Raffelt}}, \
  and\ \bibinfo {author} {\bibfnamefont {E.}~\bibnamefont {Vitagliano}},\
  }\href {\doibase 10.1103/PhysRevD.101.015025} {\bibfield  {journal} {\bibinfo
   {journal} {Phys. Rev. D}\ }\textbf {\bibinfo {volume} {101}},\ \bibinfo
  {pages} {015025} (\bibinfo {year} {2020})},\ \Eprint
  {http://arxiv.org/abs/1905.01264} {arXiv:1905.01264 [hep-ph]} \BibitemShut
  {NoStop}%
\bibitem [{\citenamefont {de~Gouv\^ea}\ \emph {et~al.}(2020)\citenamefont
  {de~Gouv\^ea}, \citenamefont {Martinez-Soler},\ and\ \citenamefont
  {Sen}}]{deGouvea:2019goq}%
  \BibitemOpen
  \bibfield  {author} {\bibinfo {author} {\bibfnamefont {A.}~\bibnamefont
  {de~Gouv\^ea}}, \bibinfo {author} {\bibfnamefont {I.}~\bibnamefont
  {Martinez-Soler}}, \ and\ \bibinfo {author} {\bibfnamefont {M.}~\bibnamefont
  {Sen}},\ }\href {\doibase 10.1103/PhysRevD.101.043013} {\bibfield  {journal}
  {\bibinfo  {journal} {Phys. Rev. D}\ }\textbf {\bibinfo {volume} {101}},\
  \bibinfo {pages} {043013} (\bibinfo {year} {2020})},\ \Eprint
  {http://arxiv.org/abs/1910.01127} {arXiv:1910.01127 [hep-ph]} \BibitemShut
  {NoStop}%
\bibitem [{\citenamefont {de~Gouvea}\ \emph {et~al.}(2021)\citenamefont
  {de~Gouvea}, \citenamefont {Fox}, \citenamefont {Kayser},\ and\ \citenamefont
  {Kelly}}]{deGouvea:2021ual}%
  \BibitemOpen
  \bibfield  {author} {\bibinfo {author} {\bibfnamefont {A.}~\bibnamefont
  {de~Gouvea}}, \bibinfo {author} {\bibfnamefont {P.~J.}\ \bibnamefont {Fox}},
  \bibinfo {author} {\bibfnamefont {B.~J.}\ \bibnamefont {Kayser}}, \ and\
  \bibinfo {author} {\bibfnamefont {K.~J.}\ \bibnamefont {Kelly}},\ }\href@noop
  {} {\  (\bibinfo {year} {2021})},\ \Eprint {http://arxiv.org/abs/2104.05719}
  {arXiv:2104.05719 [hep-ph]} \BibitemShut {NoStop}%
\bibitem [{\citenamefont {Wolfenstein}(1981)}]{Wolfenstein:1981kw}%
  \BibitemOpen
  \bibfield  {author} {\bibinfo {author} {\bibfnamefont {L.}~\bibnamefont
  {Wolfenstein}},\ }\href {\doibase 10.1016/0550-3213(81)90096-1} {\bibfield
  {journal} {\bibinfo  {journal} {Nucl. Phys.}\ }\textbf {\bibinfo {volume}
  {B186}},\ \bibinfo {pages} {147} (\bibinfo {year} {1981})}\BibitemShut
  {NoStop}%
%%CITATION = NUPHA,B186,147;%%
\bibitem [{\citenamefont {Petcov}(1982)}]{Petcov:1982ya}%
  \BibitemOpen
  \bibfield  {author} {\bibinfo {author} {\bibfnamefont {S.~T.}\ \bibnamefont
  {Petcov}},\ }\href {\doibase 10.1016/0370-2693(82)91246-1} {\bibfield
  {journal} {\bibinfo  {journal} {Phys. Lett.}\ }\textbf {\bibinfo {volume}
  {110B}},\ \bibinfo {pages} {245} (\bibinfo {year} {1982})}\BibitemShut
  {NoStop}%
%%CITATION = PHLTA,110B,245;%%
\bibitem [{\citenamefont {Bilenky}\ and\ \citenamefont
  {Pontecorvo}(1983)}]{Bilenky:1983wt}%
  \BibitemOpen
  \bibfield  {author} {\bibinfo {author} {\bibfnamefont {S.~M.}\ \bibnamefont
  {Bilenky}}\ and\ \bibinfo {author} {\bibfnamefont {B.}~\bibnamefont
  {Pontecorvo}},\ }\href@noop {} {\bibfield  {journal} {\bibinfo  {journal}
  {Sov. J. Nucl. Phys.}\ }\textbf {\bibinfo {volume} {38}},\ \bibinfo {pages}
  {248} (\bibinfo {year} {1983})},\ \bibinfo {note} {[Lett. Nuovo
  Cim.37,467(1983); Yad. Fiz.38,415(1983)]}\BibitemShut {NoStop}%
%%CITATION = SJNCA,38,248;%%
\bibitem [{\citenamefont {Kobayashi}\ and\ \citenamefont
  {Lim}(2001)}]{Kobayashi:2000md}%
  \BibitemOpen
  \bibfield  {author} {\bibinfo {author} {\bibfnamefont {M.}~\bibnamefont
  {Kobayashi}}\ and\ \bibinfo {author} {\bibfnamefont {C.~S.}\ \bibnamefont
  {Lim}},\ }\href {\doibase 10.1103/PhysRevD.64.013003} {\bibfield  {journal}
  {\bibinfo  {journal} {Phys. Rev.}\ }\textbf {\bibinfo {volume} {D64}},\
  \bibinfo {pages} {013003} (\bibinfo {year} {2001})},\ \Eprint
  {http://arxiv.org/abs/hep-ph/0012266} {arXiv:hep-ph/0012266 [hep-ph]}
  \BibitemShut {NoStop}%
%%CITATION = HEP-PH/0012266;%%
\bibitem [{\citenamefont {Anamiati}\ \emph {et~al.}(2018)\citenamefont
  {Anamiati}, \citenamefont {Fonseca},\ and\ \citenamefont
  {Hirsch}}]{Anamiati:2017rxw}%
  \BibitemOpen
  \bibfield  {author} {\bibinfo {author} {\bibfnamefont {G.}~\bibnamefont
  {Anamiati}}, \bibinfo {author} {\bibfnamefont {R.~M.}\ \bibnamefont
  {Fonseca}}, \ and\ \bibinfo {author} {\bibfnamefont {M.}~\bibnamefont
  {Hirsch}},\ }\href {\doibase 10.1103/PhysRevD.97.095008} {\bibfield
  {journal} {\bibinfo  {journal} {Phys. Rev. D}\ }\textbf {\bibinfo {volume}
  {97}},\ \bibinfo {pages} {095008} (\bibinfo {year} {2018})},\ \Eprint
  {http://arxiv.org/abs/1710.06249} {arXiv:1710.06249 [hep-ph]} \BibitemShut
  {NoStop}%
\bibitem [{\citenamefont {de~Gouvea}\ \emph {et~al.}(2009)\citenamefont
  {de~Gouvea}, \citenamefont {Huang},\ and\ \citenamefont
  {Jenkins}}]{deGouvea:2009fp}%
  \BibitemOpen
  \bibfield  {author} {\bibinfo {author} {\bibfnamefont {A.}~\bibnamefont
  {de~Gouvea}}, \bibinfo {author} {\bibfnamefont {W.-C.}\ \bibnamefont
  {Huang}}, \ and\ \bibinfo {author} {\bibfnamefont {J.}~\bibnamefont
  {Jenkins}},\ }\href {\doibase 10.1103/PhysRevD.80.073007} {\bibfield
  {journal} {\bibinfo  {journal} {Phys. Rev.}\ }\textbf {\bibinfo {volume}
  {D80}},\ \bibinfo {pages} {073007} (\bibinfo {year} {2009})},\ \Eprint
  {http://arxiv.org/abs/0906.1611} {arXiv:0906.1611 [hep-ph]} \BibitemShut
  {NoStop}%
%%CITATION = ARXIV:0906.1611;%%
\bibitem [{\citenamefont {Vissani}\ and\ \citenamefont
  {Boeltzig}(2015)}]{Vissani:2015pss}%
  \BibitemOpen
  \bibfield  {author} {\bibinfo {author} {\bibfnamefont {F.}~\bibnamefont
  {Vissani}}\ and\ \bibinfo {author} {\bibfnamefont {A.}~\bibnamefont
  {Boeltzig}},\ }\href {\doibase 10.22323/1.244.0008} {\bibfield  {journal}
  {\bibinfo  {journal} {PoS}\ }\textbf {\bibinfo {volume} {NEUTEL2015}},\
  \bibinfo {pages} {008} (\bibinfo {year} {2015})}\BibitemShut {NoStop}%
\bibitem [{\citenamefont {Berezinsky}\ \emph {et~al.}(2003)\citenamefont
  {Berezinsky}, \citenamefont {Narayan},\ and\ \citenamefont
  {Vissani}}]{Berezinsky:2002fa}%
  \BibitemOpen
  \bibfield  {author} {\bibinfo {author} {\bibfnamefont {V.}~\bibnamefont
  {Berezinsky}}, \bibinfo {author} {\bibfnamefont {M.}~\bibnamefont {Narayan}},
  \ and\ \bibinfo {author} {\bibfnamefont {F.}~\bibnamefont {Vissani}},\ }\href
  {\doibase 10.1016/S0550-3213(03)00191-3} {\bibfield  {journal} {\bibinfo
  {journal} {Nucl. Phys. B}\ }\textbf {\bibinfo {volume} {658}},\ \bibinfo
  {pages} {254} (\bibinfo {year} {2003})},\ \Eprint
  {http://arxiv.org/abs/hep-ph/0210204} {arXiv:hep-ph/0210204} \BibitemShut
  {NoStop}%
\bibitem [{\citenamefont {Esteban}\ \emph {et~al.}(2020)\citenamefont
  {Esteban}, \citenamefont {Gonzalez-Garcia}, \citenamefont {Maltoni},
  \citenamefont {Schwetz},\ and\ \citenamefont {Zhou}}]{Esteban:2020cvm}%
  \BibitemOpen
  \bibfield  {author} {\bibinfo {author} {\bibfnamefont {I.}~\bibnamefont
  {Esteban}}, \bibinfo {author} {\bibfnamefont {M.~C.}\ \bibnamefont
  {Gonzalez-Garcia}}, \bibinfo {author} {\bibfnamefont {M.}~\bibnamefont
  {Maltoni}}, \bibinfo {author} {\bibfnamefont {T.}~\bibnamefont {Schwetz}}, \
  and\ \bibinfo {author} {\bibfnamefont {A.}~\bibnamefont {Zhou}},\ }\href
  {\doibase 10.1007/JHEP09(2020)178} {\bibfield  {journal} {\bibinfo  {journal}
  {JHEP}\ }\textbf {\bibinfo {volume} {09}},\ \bibinfo {pages} {178} (\bibinfo
  {year} {2020})},\ \Eprint {http://arxiv.org/abs/2007.14792} {arXiv:2007.14792
  [hep-ph]} \BibitemShut {NoStop}%
\bibitem [{\citenamefont {Beacom}\ \emph {et~al.}(2004)\citenamefont {Beacom},
  \citenamefont {Bell}, \citenamefont {Hooper}, \citenamefont {Learned},
  \citenamefont {Pakvasa},\ and\ \citenamefont {Weiler}}]{Beacom:2003eu}%
  \BibitemOpen
  \bibfield  {author} {\bibinfo {author} {\bibfnamefont {J.~F.}\ \bibnamefont
  {Beacom}}, \bibinfo {author} {\bibfnamefont {N.~F.}\ \bibnamefont {Bell}},
  \bibinfo {author} {\bibfnamefont {D.}~\bibnamefont {Hooper}}, \bibinfo
  {author} {\bibfnamefont {J.~G.}\ \bibnamefont {Learned}}, \bibinfo {author}
  {\bibfnamefont {S.}~\bibnamefont {Pakvasa}}, \ and\ \bibinfo {author}
  {\bibfnamefont {T.~J.}\ \bibnamefont {Weiler}},\ }\href {\doibase
  10.1103/PhysRevLett.92.011101} {\bibfield  {journal} {\bibinfo  {journal}
  {Phys. Rev. Lett.}\ }\textbf {\bibinfo {volume} {92}},\ \bibinfo {pages}
  {011101} (\bibinfo {year} {2004})},\ \Eprint
  {http://arxiv.org/abs/hep-ph/0307151} {arXiv:hep-ph/0307151 [hep-ph]}
  \BibitemShut {NoStop}%
%%CITATION = HEP-PH/0307151;%%
\bibitem [{\citenamefont {De~Gouv\^ea}\ \emph {et~al.}(2020)\citenamefont
  {De~Gouv\^ea}, \citenamefont {Martinez-Soler}, \citenamefont
  {Perez-Gonzalez},\ and\ \citenamefont {Sen}}]{deGouvea:2020eqq}%
  \BibitemOpen
  \bibfield  {author} {\bibinfo {author} {\bibfnamefont {A.}~\bibnamefont
  {De~Gouv\^ea}}, \bibinfo {author} {\bibfnamefont {I.}~\bibnamefont
  {Martinez-Soler}}, \bibinfo {author} {\bibfnamefont {Y.~F.}\ \bibnamefont
  {Perez-Gonzalez}}, \ and\ \bibinfo {author} {\bibfnamefont {M.}~\bibnamefont
  {Sen}},\ }\href {\doibase 10.1103/PhysRevD.102.123012} {\bibfield  {journal}
  {\bibinfo  {journal} {Phys. Rev. D}\ }\textbf {\bibinfo {volume} {102}},\
  \bibinfo {pages} {123012} (\bibinfo {year} {2020})},\ \Eprint
  {http://arxiv.org/abs/2007.13748} {arXiv:2007.13748 [hep-ph]} \BibitemShut
  {NoStop}%
\bibitem [{\citenamefont {Anamiati}\ \emph {et~al.}(2019)\citenamefont
  {Anamiati}, \citenamefont {De~Romeri}, \citenamefont {Hirsch}, \citenamefont
  {Ternes},\ and\ \citenamefont {T\'ortola}}]{Anamiati:2019maf}%
  \BibitemOpen
  \bibfield  {author} {\bibinfo {author} {\bibfnamefont {G.}~\bibnamefont
  {Anamiati}}, \bibinfo {author} {\bibfnamefont {V.}~\bibnamefont {De~Romeri}},
  \bibinfo {author} {\bibfnamefont {M.}~\bibnamefont {Hirsch}}, \bibinfo
  {author} {\bibfnamefont {C.~A.}\ \bibnamefont {Ternes}}, \ and\ \bibinfo
  {author} {\bibfnamefont {M.}~\bibnamefont {T\'ortola}},\ }\href {\doibase
  10.1103/PhysRevD.100.035032} {\bibfield  {journal} {\bibinfo  {journal}
  {Phys. Rev. D}\ }\textbf {\bibinfo {volume} {100}},\ \bibinfo {pages}
  {035032} (\bibinfo {year} {2019})},\ \Eprint
  {http://arxiv.org/abs/1907.00980} {arXiv:1907.00980 [hep-ph]} \BibitemShut
  {NoStop}%
\bibitem [{\citenamefont {Das}\ \emph {et~al.}(2014)\citenamefont {Das},
  \citenamefont {Bhupal~Dev},\ and\ \citenamefont {Okada}}]{Das:2014jxa}%
  \BibitemOpen
  \bibfield  {author} {\bibinfo {author} {\bibfnamefont {A.}~\bibnamefont
  {Das}}, \bibinfo {author} {\bibfnamefont {P.~S.}\ \bibnamefont {Bhupal~Dev}},
  \ and\ \bibinfo {author} {\bibfnamefont {N.}~\bibnamefont {Okada}},\ }\href
  {\doibase 10.1016/j.physletb.2014.06.058} {\bibfield  {journal} {\bibinfo
  {journal} {Phys. Lett. B}\ }\textbf {\bibinfo {volume} {735}},\ \bibinfo
  {pages} {364} (\bibinfo {year} {2014})},\ \Eprint
  {http://arxiv.org/abs/1405.0177} {arXiv:1405.0177 [hep-ph]} \BibitemShut
  {NoStop}%
\bibitem [{\citenamefont {Hern\'andez}\ \emph {et~al.}(2019)\citenamefont
  {Hern\'andez}, \citenamefont {Jones-P\'erez},\ and\ \citenamefont
  {Suarez-Navarro}}]{Hernandez:2018cgc}%
  \BibitemOpen
  \bibfield  {author} {\bibinfo {author} {\bibfnamefont {P.}~\bibnamefont
  {Hern\'andez}}, \bibinfo {author} {\bibfnamefont {J.}~\bibnamefont
  {Jones-P\'erez}}, \ and\ \bibinfo {author} {\bibfnamefont {O.}~\bibnamefont
  {Suarez-Navarro}},\ }\href {\doibase 10.1140/epjc/s10052-019-6728-1}
  {\bibfield  {journal} {\bibinfo  {journal} {Eur. Phys. J. C}\ }\textbf
  {\bibinfo {volume} {79}},\ \bibinfo {pages} {220} (\bibinfo {year} {2019})},\
  \Eprint {http://arxiv.org/abs/1810.07210} {arXiv:1810.07210 [hep-ph]}
  \BibitemShut {NoStop}%
\bibitem [{\citenamefont {Keranen}\ \emph {et~al.}(2003)\citenamefont
  {Keranen}, \citenamefont {Maalampi}, \citenamefont {Myyrylainen},\ and\
  \citenamefont {Riittinen}}]{Keranen:2003xd}%
  \BibitemOpen
  \bibfield  {author} {\bibinfo {author} {\bibfnamefont {P.}~\bibnamefont
  {Keranen}}, \bibinfo {author} {\bibfnamefont {J.}~\bibnamefont {Maalampi}},
  \bibinfo {author} {\bibfnamefont {M.}~\bibnamefont {Myyrylainen}}, \ and\
  \bibinfo {author} {\bibfnamefont {J.}~\bibnamefont {Riittinen}},\ }\href
  {\doibase 10.1016/j.physletb.2003.09.006} {\bibfield  {journal} {\bibinfo
  {journal} {Phys. Lett. B}\ }\textbf {\bibinfo {volume} {574}},\ \bibinfo
  {pages} {162} (\bibinfo {year} {2003})},\ \Eprint
  {http://arxiv.org/abs/hep-ph/0307041} {arXiv:hep-ph/0307041} \BibitemShut
  {NoStop}%
\bibitem [{\citenamefont {Esmaili}(2010)}]{Esmaili:2009fk}%
  \BibitemOpen
  \bibfield  {author} {\bibinfo {author} {\bibfnamefont {A.}~\bibnamefont
  {Esmaili}},\ }\href {\doibase 10.1103/PhysRevD.81.013006} {\bibfield
  {journal} {\bibinfo  {journal} {Phys. Rev. D}\ }\textbf {\bibinfo {volume}
  {81}},\ \bibinfo {pages} {013006} (\bibinfo {year} {2010})},\ \Eprint
  {http://arxiv.org/abs/0909.5410} {arXiv:0909.5410 [hep-ph]} \BibitemShut
  {NoStop}%
\bibitem [{\citenamefont {Esmaili}\ and\ \citenamefont
  {Farzan}(2012)}]{Esmaili:2012ac}%
  \BibitemOpen
  \bibfield  {author} {\bibinfo {author} {\bibfnamefont {A.}~\bibnamefont
  {Esmaili}}\ and\ \bibinfo {author} {\bibfnamefont {Y.}~\bibnamefont
  {Farzan}},\ }\href {\doibase 10.1088/1475-7516/2012/12/014} {\bibfield
  {journal} {\bibinfo  {journal} {JCAP}\ }\textbf {\bibinfo {volume} {1212}},\
  \bibinfo {pages} {014} (\bibinfo {year} {2012})},\ \Eprint
  {http://arxiv.org/abs/1208.6012} {arXiv:1208.6012 [hep-ph]} \BibitemShut
  {NoStop}%
%%CITATION = ARXIV:1208.6012;%%
\bibitem [{\citenamefont {Joshipura}\ \emph {et~al.}(2014)\citenamefont
  {Joshipura}, \citenamefont {Mohanty},\ and\ \citenamefont
  {Pakvasa}}]{Joshipura:2013yba}%
  \BibitemOpen
  \bibfield  {author} {\bibinfo {author} {\bibfnamefont {A.~S.}\ \bibnamefont
  {Joshipura}}, \bibinfo {author} {\bibfnamefont {S.}~\bibnamefont {Mohanty}},
  \ and\ \bibinfo {author} {\bibfnamefont {S.}~\bibnamefont {Pakvasa}},\ }\href
  {\doibase 10.1103/PhysRevD.89.033003} {\bibfield  {journal} {\bibinfo
  {journal} {Phys. Rev. D}\ }\textbf {\bibinfo {volume} {89}},\ \bibinfo
  {pages} {033003} (\bibinfo {year} {2014})},\ \Eprint
  {http://arxiv.org/abs/1307.5712} {arXiv:1307.5712 [hep-ph]} \BibitemShut
  {NoStop}%
\bibitem [{\citenamefont {Brdar}\ and\ \citenamefont
  {Hansen}(2019)}]{Brdar:2018tce}%
  \BibitemOpen
  \bibfield  {author} {\bibinfo {author} {\bibfnamefont {V.}~\bibnamefont
  {Brdar}}\ and\ \bibinfo {author} {\bibfnamefont {R.~S.~L.}\ \bibnamefont
  {Hansen}},\ }\href {\doibase 10.1088/1475-7516/2019/02/023} {\bibfield
  {journal} {\bibinfo  {journal} {JCAP}\ }\textbf {\bibinfo {volume} {02}},\
  \bibinfo {pages} {023} (\bibinfo {year} {2019})},\ \Eprint
  {http://arxiv.org/abs/1812.05541} {arXiv:1812.05541 [hep-ph]} \BibitemShut
  {NoStop}%
\bibitem [{\citenamefont {Hirata}\ \emph {et~al.}(1987)\citenamefont {Hirata}
  \emph {et~al.}}]{Hirata:1987hu}%
  \BibitemOpen
  \bibfield  {author} {\bibinfo {author} {\bibfnamefont {K.}~\bibnamefont
  {Hirata}} \emph {et~al.} (\bibinfo {collaboration} {Kamiokande-II}),\ }\href
  {\doibase 10.1103/PhysRevLett.58.1490} {\bibfield  {journal} {\bibinfo
  {journal} {Phys. Rev. Lett.}\ }\textbf {\bibinfo {volume} {58}},\ \bibinfo
  {pages} {1490} (\bibinfo {year} {1987})}\BibitemShut {NoStop}%
\bibitem [{\citenamefont {Hirata}\ \emph {et~al.}(1988)\citenamefont {Hirata}
  \emph {et~al.}}]{Hirata:1988ad}%
  \BibitemOpen
  \bibfield  {author} {\bibinfo {author} {\bibfnamefont {K.~S.}\ \bibnamefont
  {Hirata}} \emph {et~al.},\ }\href {\doibase 10.1103/PhysRevD.38.448}
  {\bibfield  {journal} {\bibinfo  {journal} {Phys. Rev. D}\ }\textbf {\bibinfo
  {volume} {38}},\ \bibinfo {pages} {448} (\bibinfo {year} {1988})}\BibitemShut
  {NoStop}%
\bibitem [{\citenamefont {Bionta}\ \emph {et~al.}(1987)\citenamefont {Bionta}
  \emph {et~al.}}]{Bionta:1987qt}%
  \BibitemOpen
  \bibfield  {author} {\bibinfo {author} {\bibfnamefont {R.~M.}\ \bibnamefont
  {Bionta}} \emph {et~al.},\ }\href {\doibase 10.1103/PhysRevLett.58.1494}
  {\bibfield  {journal} {\bibinfo  {journal} {Phys. Rev. Lett.}\ }\textbf
  {\bibinfo {volume} {58}},\ \bibinfo {pages} {1494} (\bibinfo {year}
  {1987})}\BibitemShut {NoStop}%
\bibitem [{\citenamefont {Bratton}\ \emph {et~al.}(1988)\citenamefont {Bratton}
  \emph {et~al.}}]{Bratton:1988ww}%
  \BibitemOpen
  \bibfield  {author} {\bibinfo {author} {\bibfnamefont {C.~B.}\ \bibnamefont
  {Bratton}} \emph {et~al.} (\bibinfo {collaboration} {IMB}),\ }\href {\doibase
  10.1103/PhysRevD.37.3361} {\bibfield  {journal} {\bibinfo  {journal} {Phys.
  Rev. D}\ }\textbf {\bibinfo {volume} {37}},\ \bibinfo {pages} {3361}
  (\bibinfo {year} {1988})}\BibitemShut {NoStop}%
\bibitem [{\citenamefont {Alekseev}\ \emph {et~al.}(1988)\citenamefont
  {Alekseev}, \citenamefont {Alekseeva}, \citenamefont {Krivosheina},\ and\
  \citenamefont {Volchenko}}]{Alekseev:1988gp}%
  \BibitemOpen
  \bibfield  {author} {\bibinfo {author} {\bibfnamefont {E.~N.}\ \bibnamefont
  {Alekseev}}, \bibinfo {author} {\bibfnamefont {L.~N.}\ \bibnamefont
  {Alekseeva}}, \bibinfo {author} {\bibfnamefont {I.~V.}\ \bibnamefont
  {Krivosheina}}, \ and\ \bibinfo {author} {\bibfnamefont {V.~I.}\ \bibnamefont
  {Volchenko}},\ }\href {\doibase 10.1016/0370-2693(88)91651-6} {\bibfield
  {journal} {\bibinfo  {journal} {Phys. Lett. B}\ }\textbf {\bibinfo {volume}
  {205}},\ \bibinfo {pages} {209} (\bibinfo {year} {1988})}\BibitemShut
  {NoStop}%
\bibitem [{\citenamefont {Mohapatra}\ and\ \citenamefont
  {Senjanovic}(1980)}]{Mohapatra:1979ia}%
  \BibitemOpen
  \bibfield  {author} {\bibinfo {author} {\bibfnamefont {R.~N.}\ \bibnamefont
  {Mohapatra}}\ and\ \bibinfo {author} {\bibfnamefont {G.}~\bibnamefont
  {Senjanovic}},\ }\href {\doibase 10.1103/PhysRevLett.44.912} {\bibfield
  {journal} {\bibinfo  {journal} {Phys. Rev. Lett.}\ }\textbf {\bibinfo
  {volume} {44}},\ \bibinfo {pages} {912} (\bibinfo {year} {1980})}\BibitemShut
  {NoStop}%
%%CITATION = PRLTA,44,912;%%
\bibitem [{\citenamefont {Gell-Mann}\ \emph {et~al.}(1979)\citenamefont
  {Gell-Mann}, \citenamefont {Ramond},\ and\ \citenamefont
  {Slansky}}]{GellMann:1980vs}%
  \BibitemOpen
  \bibfield  {author} {\bibinfo {author} {\bibfnamefont {M.}~\bibnamefont
  {Gell-Mann}}, \bibinfo {author} {\bibfnamefont {P.}~\bibnamefont {Ramond}}, \
  and\ \bibinfo {author} {\bibfnamefont {R.}~\bibnamefont {Slansky}},\
  }\bibfield  {booktitle} {\emph {\bibinfo {booktitle} {{Supergravity Workshop
  Stony Brook}}},\ }\href@noop {} {\bibfield  {journal} {\bibinfo  {journal}
  {Conf. Proc.}\ }\textbf {\bibinfo {volume} {C790927}},\ \bibinfo {pages}
  {315} (\bibinfo {year} {1979})},\ \Eprint {http://arxiv.org/abs/1306.4669}
  {arXiv:1306.4669 [hep-th]} \BibitemShut {NoStop}%
%%CITATION = ARXIV:1306.4669;%%
\bibitem [{\citenamefont {Yanagida}(1979)}]{Yanagida:1979as}%
  \BibitemOpen
  \bibfield  {author} {\bibinfo {author} {\bibfnamefont {T.}~\bibnamefont
  {Yanagida}},\ }\href@noop {} {\bibfield  {journal} {\bibinfo  {journal}
  {Conf. Proc.}\ }\textbf {\bibinfo {volume} {C7902131}},\ \bibinfo {pages}
  {95} (\bibinfo {year} {1979})}\BibitemShut {NoStop}%
%%CITATION = CONFP,C7902131,95;%%
\bibitem [{\citenamefont {Minkowski}(1977)}]{Minkowski:1977sc}%
  \BibitemOpen
  \bibfield  {author} {\bibinfo {author} {\bibfnamefont {P.}~\bibnamefont
  {Minkowski}},\ }\href {\doibase 10.1016/0370-2693(77)90435-X} {\bibfield
  {journal} {\bibinfo  {journal} {Phys. Lett.}\ }\textbf {\bibinfo {volume}
  {67B}},\ \bibinfo {pages} {421} (\bibinfo {year} {1977})}\BibitemShut
  {NoStop}%
%%CITATION = PHLTA,67B,421;%%
\bibitem [{\citenamefont {Mohapatra}\ and\ \citenamefont
  {Senjanovic}(1981)}]{Mohapatra:1980yp}%
  \BibitemOpen
  \bibfield  {author} {\bibinfo {author} {\bibfnamefont {R.~N.}\ \bibnamefont
  {Mohapatra}}\ and\ \bibinfo {author} {\bibfnamefont {G.}~\bibnamefont
  {Senjanovic}},\ }\href {\doibase 10.1103/PhysRevD.23.165} {\bibfield
  {journal} {\bibinfo  {journal} {Phys. Rev.}\ }\textbf {\bibinfo {volume}
  {D23}},\ \bibinfo {pages} {165} (\bibinfo {year} {1981})}\BibitemShut
  {NoStop}%
%%CITATION = PHRVA,D23,165;%%
\bibitem [{\citenamefont {Magg}\ and\ \citenamefont
  {Wetterich}(1980)}]{Magg:1980ut}%
  \BibitemOpen
  \bibfield  {author} {\bibinfo {author} {\bibfnamefont {M.}~\bibnamefont
  {Magg}}\ and\ \bibinfo {author} {\bibfnamefont {C.}~\bibnamefont
  {Wetterich}},\ }\href {\doibase 10.1016/0370-2693(80)90825-4} {\bibfield
  {journal} {\bibinfo  {journal} {Phys. Lett.}\ }\textbf {\bibinfo {volume}
  {B94}},\ \bibinfo {pages} {61} (\bibinfo {year} {1980})}\BibitemShut
  {NoStop}%
%%CITATION = PHLTA,B94,61;%%
\bibitem [{\citenamefont {Lazarides}\ \emph {et~al.}(1981)\citenamefont
  {Lazarides}, \citenamefont {Shafi},\ and\ \citenamefont
  {Wetterich}}]{Lazarides:1980nt}%
  \BibitemOpen
  \bibfield  {author} {\bibinfo {author} {\bibfnamefont {G.}~\bibnamefont
  {Lazarides}}, \bibinfo {author} {\bibfnamefont {Q.}~\bibnamefont {Shafi}}, \
  and\ \bibinfo {author} {\bibfnamefont {C.}~\bibnamefont {Wetterich}},\ }\href
  {\doibase 10.1016/0550-3213(81)90354-0} {\bibfield  {journal} {\bibinfo
  {journal} {Nucl. Phys.}\ }\textbf {\bibinfo {volume} {B181}},\ \bibinfo
  {pages} {287} (\bibinfo {year} {1981})}\BibitemShut {NoStop}%
%%CITATION = NUPHA,B181,287;%%
\bibitem [{\citenamefont {Wetterich}(1981)}]{Wetterich:1981bx}%
  \BibitemOpen
  \bibfield  {author} {\bibinfo {author} {\bibfnamefont {C.}~\bibnamefont
  {Wetterich}},\ }\href {\doibase 10.1016/0550-3213(81)90279-0} {\bibfield
  {journal} {\bibinfo  {journal} {Nucl. Phys.}\ }\textbf {\bibinfo {volume}
  {B187}},\ \bibinfo {pages} {343} (\bibinfo {year} {1981})}\BibitemShut
  {NoStop}%
%%CITATION = NUPHA,B187,343;%%
\bibitem [{\citenamefont {Porto-Silva}\ and\ \citenamefont
  {Smirnov}(2021)}]{Porto-Silva:2021ael}%
  \BibitemOpen
  \bibfield  {author} {\bibinfo {author} {\bibfnamefont {Y.~P.}\ \bibnamefont
  {Porto-Silva}}\ and\ \bibinfo {author} {\bibfnamefont {A.~Y.}\ \bibnamefont
  {Smirnov}},\ }\href@noop {} {\  (\bibinfo {year} {2021})},\ \Eprint
  {http://arxiv.org/abs/2103.10149} {arXiv:2103.10149 [hep-ph]} \BibitemShut
  {NoStop}%
\bibitem [{\citenamefont {Kersten}\ and\ \citenamefont
  {Smirnov}(2016)}]{Kersten:2015kio}%
  \BibitemOpen
  \bibfield  {author} {\bibinfo {author} {\bibfnamefont {J.}~\bibnamefont
  {Kersten}}\ and\ \bibinfo {author} {\bibfnamefont {A.~Y.}\ \bibnamefont
  {Smirnov}},\ }\href {\doibase 10.1140/epjc/s10052-016-4187-5} {\bibfield
  {journal} {\bibinfo  {journal} {Eur. Phys. J. C}\ }\textbf {\bibinfo {volume}
  {76}},\ \bibinfo {pages} {339} (\bibinfo {year} {2016})},\ \Eprint
  {http://arxiv.org/abs/1512.09068} {arXiv:1512.09068 [hep-ph]} \BibitemShut
  {NoStop}%
\bibitem [{\citenamefont {Tamborra}\ \emph {et~al.}(2012)\citenamefont
  {Tamborra}, \citenamefont {Muller}, \citenamefont {Hudepohl}, \citenamefont
  {Janka},\ and\ \citenamefont {Raffelt}}]{Tamborra:2012ac}%
  \BibitemOpen
  \bibfield  {author} {\bibinfo {author} {\bibfnamefont {I.}~\bibnamefont
  {Tamborra}}, \bibinfo {author} {\bibfnamefont {B.}~\bibnamefont {Muller}},
  \bibinfo {author} {\bibfnamefont {L.}~\bibnamefont {Hudepohl}}, \bibinfo
  {author} {\bibfnamefont {H.-T.}\ \bibnamefont {Janka}}, \ and\ \bibinfo
  {author} {\bibfnamefont {G.}~\bibnamefont {Raffelt}},\ }\href {\doibase
  10.1103/PhysRevD.86.125031} {\bibfield  {journal} {\bibinfo  {journal} {Phys.
  Rev. D}\ }\textbf {\bibinfo {volume} {86}},\ \bibinfo {pages} {125031}
  (\bibinfo {year} {2012})},\ \Eprint {http://arxiv.org/abs/1211.3920}
  {arXiv:1211.3920 [astro-ph.SR]} \BibitemShut {NoStop}%
\bibitem [{\citenamefont {Lunardini}(2006)}]{Lunardini:2005jf}%
  \BibitemOpen
  \bibfield  {author} {\bibinfo {author} {\bibfnamefont {C.}~\bibnamefont
  {Lunardini}},\ }\href {\doibase 10.1016/j.astropartphys.2006.06.008}
  {\bibfield  {journal} {\bibinfo  {journal} {Astropart. Phys.}\ }\textbf
  {\bibinfo {volume} {26}},\ \bibinfo {pages} {190} (\bibinfo {year} {2006})},\
  \Eprint {http://arxiv.org/abs/astro-ph/0509233} {arXiv:astro-ph/0509233}
  \BibitemShut {NoStop}%
\bibitem [{\citenamefont {Wolfenstein}(1978)}]{PhysRevD.17.2369}%
  \BibitemOpen
  \bibfield  {author} {\bibinfo {author} {\bibfnamefont {L.}~\bibnamefont
  {Wolfenstein}},\ }\href {\doibase 10.1103/PhysRevD.17.2369} {\bibfield
  {journal} {\bibinfo  {journal} {Phys. Rev. D}\ }\textbf {\bibinfo {volume}
  {17}},\ \bibinfo {pages} {2369} (\bibinfo {year} {1978})}\BibitemShut
  {NoStop}%
\bibitem [{\citenamefont {Mikheev}\ and\ \citenamefont
  {Smirnov}(1985)}]{Mikheev:1986gs}%
  \BibitemOpen
  \bibfield  {author} {\bibinfo {author} {\bibfnamefont {S.~P.}\ \bibnamefont
  {Mikheev}}\ and\ \bibinfo {author} {\bibfnamefont {A.~{\relax Yu}.}\
  \bibnamefont {Smirnov}},\ }\href@noop {} {\bibfield  {journal} {\bibinfo
  {journal} {Sov. J. Nucl. Phys.}\ }\textbf {\bibinfo {volume} {42}},\ \bibinfo
  {pages} {913} (\bibinfo {year} {1985})},\ \bibinfo {note} {[Yad.
  Fiz.42,1441(1985)]}\BibitemShut {NoStop}%
%%CITATION = SJNCA,42,913;%%
\bibitem [{\citenamefont {Dighe}\ and\ \citenamefont
  {Smirnov}(2000)}]{Dighe:1999bi}%
  \BibitemOpen
  \bibfield  {author} {\bibinfo {author} {\bibfnamefont {A.~S.}\ \bibnamefont
  {Dighe}}\ and\ \bibinfo {author} {\bibfnamefont {A.~{\relax Yu}.}\
  \bibnamefont {Smirnov}},\ }\href {\doibase 10.1103/PhysRevD.62.033007}
  {\bibfield  {journal} {\bibinfo  {journal} {Phys. Rev.}\ }\textbf {\bibinfo
  {volume} {D62}},\ \bibinfo {pages} {033007} (\bibinfo {year} {2000})},\
  \Eprint {http://arxiv.org/abs/hep-ph/9907423} {arXiv:hep-ph/9907423 [hep-ph]}
  \BibitemShut {NoStop}%
%%CITATION = HEP-PH/9907423;%%
\bibitem [{\citenamefont {Keranen}\ \emph {et~al.}(2004)\citenamefont
  {Keranen}, \citenamefont {Maalampi}, \citenamefont {Myyrylainen},\ and\
  \citenamefont {Riittinen}}]{Keranen:2004rg}%
  \BibitemOpen
  \bibfield  {author} {\bibinfo {author} {\bibfnamefont {P.}~\bibnamefont
  {Keranen}}, \bibinfo {author} {\bibfnamefont {J.}~\bibnamefont {Maalampi}},
  \bibinfo {author} {\bibfnamefont {M.}~\bibnamefont {Myyrylainen}}, \ and\
  \bibinfo {author} {\bibfnamefont {J.}~\bibnamefont {Riittinen}},\ }\href
  {\doibase 10.1016/j.physletb.2004.07.041} {\bibfield  {journal} {\bibinfo
  {journal} {Phys. Lett. B}\ }\textbf {\bibinfo {volume} {597}},\ \bibinfo
  {pages} {374} (\bibinfo {year} {2004})},\ \Eprint
  {http://arxiv.org/abs/hep-ph/0401082} {arXiv:hep-ph/0401082} \BibitemShut
  {NoStop}%
\bibitem [{\citenamefont {Vissani}(2015)}]{Vissani:2014doa}%
  \BibitemOpen
  \bibfield  {author} {\bibinfo {author} {\bibfnamefont {F.}~\bibnamefont
  {Vissani}},\ }\href {\doibase 10.1088/0954-3899/42/1/013001} {\bibfield
  {journal} {\bibinfo  {journal} {J. Phys. G}\ }\textbf {\bibinfo {volume}
  {42}},\ \bibinfo {pages} {013001} (\bibinfo {year} {2015})},\ \Eprint
  {http://arxiv.org/abs/1409.4710} {arXiv:1409.4710 [astro-ph.HE]} \BibitemShut
  {NoStop}%
\bibitem [{\citenamefont {Jegerlehner}\ \emph {et~al.}(1996)\citenamefont
  {Jegerlehner}, \citenamefont {Neubig},\ and\ \citenamefont
  {Raffelt}}]{Jegerlehner:1996kx}%
  \BibitemOpen
  \bibfield  {author} {\bibinfo {author} {\bibfnamefont {B.}~\bibnamefont
  {Jegerlehner}}, \bibinfo {author} {\bibfnamefont {F.}~\bibnamefont {Neubig}},
  \ and\ \bibinfo {author} {\bibfnamefont {G.}~\bibnamefont {Raffelt}},\ }\href
  {\doibase 10.1103/PhysRevD.54.1194} {\bibfield  {journal} {\bibinfo
  {journal} {Phys. Rev. D}\ }\textbf {\bibinfo {volume} {54}},\ \bibinfo
  {pages} {1194} (\bibinfo {year} {1996})},\ \Eprint
  {http://arxiv.org/abs/astro-ph/9601111} {arXiv:astro-ph/9601111} \BibitemShut
  {NoStop}%
\bibitem [{\citenamefont {Mirizzi}\ and\ \citenamefont
  {Raffelt}(2005)}]{Mirizzi:2005tg}%
  \BibitemOpen
  \bibfield  {author} {\bibinfo {author} {\bibfnamefont {A.}~\bibnamefont
  {Mirizzi}}\ and\ \bibinfo {author} {\bibfnamefont {G.~G.}\ \bibnamefont
  {Raffelt}},\ }\href {\doibase 10.1103/PhysRevD.72.063001} {\bibfield
  {journal} {\bibinfo  {journal} {Phys. Rev. D}\ }\textbf {\bibinfo {volume}
  {72}},\ \bibinfo {pages} {063001} (\bibinfo {year} {2005})},\ \Eprint
  {http://arxiv.org/abs/astro-ph/0508612} {arXiv:astro-ph/0508612} \BibitemShut
  {NoStop}%
\bibitem [{\citenamefont {Ianni}\ \emph {et~al.}(2009)\citenamefont {Ianni},
  \citenamefont {Pagliaroli}, \citenamefont {Strumia}, \citenamefont {Torres},
  \citenamefont {Villante},\ and\ \citenamefont {Vissani}}]{Ianni:2009bd}%
  \BibitemOpen
  \bibfield  {author} {\bibinfo {author} {\bibfnamefont {A.}~\bibnamefont
  {Ianni}}, \bibinfo {author} {\bibfnamefont {G.}~\bibnamefont {Pagliaroli}},
  \bibinfo {author} {\bibfnamefont {A.}~\bibnamefont {Strumia}}, \bibinfo
  {author} {\bibfnamefont {F.~R.}\ \bibnamefont {Torres}}, \bibinfo {author}
  {\bibfnamefont {F.~L.}\ \bibnamefont {Villante}}, \ and\ \bibinfo {author}
  {\bibfnamefont {F.}~\bibnamefont {Vissani}},\ }\href {\doibase
  10.1103/PhysRevD.80.043007} {\bibfield  {journal} {\bibinfo  {journal} {Phys.
  Rev. D}\ }\textbf {\bibinfo {volume} {80}},\ \bibinfo {pages} {043007}
  (\bibinfo {year} {2009})},\ \Eprint {http://arxiv.org/abs/0907.1891}
  {arXiv:0907.1891 [hep-ph]} \BibitemShut {NoStop}%
\bibitem [{\citenamefont {Midorikawa}\ \emph {et~al.}(1987)\citenamefont
  {Midorikawa}, \citenamefont {Terazawa},\ and\ \citenamefont
  {Akama}}]{Midorikawa:1987kv}%
  \BibitemOpen
  \bibfield  {author} {\bibinfo {author} {\bibfnamefont {S.}~\bibnamefont
  {Midorikawa}}, \bibinfo {author} {\bibfnamefont {H.}~\bibnamefont
  {Terazawa}}, \ and\ \bibinfo {author} {\bibfnamefont {K.}~\bibnamefont
  {Akama}},\ }\href {\doibase 10.1142/S0217732387000690} {\bibfield  {journal}
  {\bibinfo  {journal} {Mod. Phys. Lett. A}\ }\textbf {\bibinfo {volume} {2}},\
  \bibinfo {pages} {561} (\bibinfo {year} {1987})}\BibitemShut {NoStop}%
\bibitem [{\citenamefont {Midorikawa}\ \emph {et~al.}(1988)\citenamefont
  {Midorikawa}, \citenamefont {Terazawa},\ and\ \citenamefont
  {Akama}}]{Midorikawa:1987rv}%
  \BibitemOpen
  \bibfield  {author} {\bibinfo {author} {\bibfnamefont {S.}~\bibnamefont
  {Midorikawa}}, \bibinfo {author} {\bibfnamefont {H.}~\bibnamefont
  {Terazawa}}, \ and\ \bibinfo {author} {\bibfnamefont {K.}~\bibnamefont
  {Akama}},\ }\href {\doibase 10.1142/S0217732388000258} {\bibfield  {journal}
  {\bibinfo  {journal} {Mod. Phys. Lett. A}\ }\textbf {\bibinfo {volume} {3}},\
  \bibinfo {pages} {215} (\bibinfo {year} {1988})},\ \bibinfo {note} {[Erratum:
  Mod.Phys.Lett.A 3, 547 (1988)]}\BibitemShut {NoStop}%
\bibitem [{\citenamefont {Abi}\ \emph {et~al.}(2020)\citenamefont {Abi} \emph
  {et~al.}}]{Abi:2020evt}%
  \BibitemOpen
  \bibfield  {author} {\bibinfo {author} {\bibfnamefont {B.}~\bibnamefont
  {Abi}} \emph {et~al.} (\bibinfo {collaboration} {DUNE}),\ }\href@noop {} {\
  (\bibinfo {year} {2020})},\ \Eprint {http://arxiv.org/abs/2002.03005}
  {arXiv:2002.03005 [hep-ex]} \BibitemShut {NoStop}%
\bibitem [{\citenamefont {Abe}\ \emph {et~al.}(2018)\citenamefont {Abe} \emph
  {et~al.}}]{Abe:2018uyc}%
  \BibitemOpen
  \bibfield  {author} {\bibinfo {author} {\bibfnamefont {K.}~\bibnamefont
  {Abe}} \emph {et~al.} (\bibinfo {collaboration} {Hyper-Kamiokande}),\
  }\href@noop {} {\  (\bibinfo {year} {2018})},\ \Eprint
  {http://arxiv.org/abs/1805.04163} {arXiv:1805.04163 [physics.ins-det]}
  \BibitemShut {NoStop}%
%%CITATION = ARXIV:1805.04163;%%
\bibitem [{\citenamefont {Gardiner}(2021)}]{Gardiner:2021qfr}%
  \BibitemOpen
  \bibfield  {author} {\bibinfo {author} {\bibfnamefont {S.}~\bibnamefont
  {Gardiner}},\ }\href@noop {} {\  (\bibinfo {year} {2021})},\ \Eprint
  {http://arxiv.org/abs/2101.11867} {arXiv:2101.11867 [nucl-th]} \BibitemShut
  {NoStop}%
\bibitem [{\citenamefont {Strumia}\ and\ \citenamefont
  {Vissani}(2003)}]{Strumia:2003zx}%
  \BibitemOpen
  \bibfield  {author} {\bibinfo {author} {\bibfnamefont {A.}~\bibnamefont
  {Strumia}}\ and\ \bibinfo {author} {\bibfnamefont {F.}~\bibnamefont
  {Vissani}},\ }\href {\doibase 10.1016/S0370-2693(03)00616-6} {\bibfield
  {journal} {\bibinfo  {journal} {Phys. Lett.}\ }\textbf {\bibinfo {volume}
  {B564}},\ \bibinfo {pages} {42} (\bibinfo {year} {2003})},\ \Eprint
  {http://arxiv.org/abs/astro-ph/0302055} {arXiv:astro-ph/0302055 [astro-ph]}
  \BibitemShut {NoStop}%
%%CITATION = ASTRO-PH/0302055;%%
\bibitem [{\citenamefont {Lang}\ \emph {et~al.}(2016)\citenamefont {Lang},
  \citenamefont {McCabe}, \citenamefont {Reichard}, \citenamefont {Selvi},\
  and\ \citenamefont {Tamborra}}]{Lang:2016zhv}%
  \BibitemOpen
  \bibfield  {author} {\bibinfo {author} {\bibfnamefont {R.~F.}\ \bibnamefont
  {Lang}}, \bibinfo {author} {\bibfnamefont {C.}~\bibnamefont {McCabe}},
  \bibinfo {author} {\bibfnamefont {S.}~\bibnamefont {Reichard}}, \bibinfo
  {author} {\bibfnamefont {M.}~\bibnamefont {Selvi}}, \ and\ \bibinfo {author}
  {\bibfnamefont {I.}~\bibnamefont {Tamborra}},\ }\href {\doibase
  10.1103/PhysRevD.94.103009} {\bibfield  {journal} {\bibinfo  {journal} {Phys.
  Rev. D}\ }\textbf {\bibinfo {volume} {94}},\ \bibinfo {pages} {103009}
  (\bibinfo {year} {2016})},\ \Eprint {http://arxiv.org/abs/1606.09243}
  {arXiv:1606.09243 [astro-ph.HE]} \BibitemShut {NoStop}%
\bibitem [{\citenamefont {Amoruso}\ \emph {et~al.}(2004)\citenamefont {Amoruso}
  \emph {et~al.}}]{Amoruso:2003sw}%
  \BibitemOpen
  \bibfield  {author} {\bibinfo {author} {\bibfnamefont {S.}~\bibnamefont
  {Amoruso}} \emph {et~al.} (\bibinfo {collaboration} {ICARUS}),\ }\href
  {\doibase 10.1140/epjc/s2004-01597-7} {\bibfield  {journal} {\bibinfo
  {journal} {Eur.\ Phys.\ J.\ C}\ }\textbf {\bibinfo {volume} {33}},\ \bibinfo
  {pages} {233} (\bibinfo {year} {2004})},\ \Eprint
  {http://arxiv.org/abs/hep-ex/0311040} {arXiv:hep-ex/0311040} \BibitemShut
  {NoStop}%
\end{thebibliography}%

\end{document}